\documentclass[twocolumn,trackchanges]{aastex701}

\usepackage{gensymb,amsmath,comment}

\begin{document}

\title{Revealing the Shiny Nature of Lava Worlds: Small Planet Phase Curves with TESS}

\author[0000-0002-0508-857X]{Brandon Park Coy}
\affiliation{Department of the Geophysical Sciences, University of Chicago, Chicago, IL, USA}
\email[show]{bpcoy@uchicago.edu}  

\author[0000-0002-0659-1783]{Michael Zhang}
\affiliation{Department of Astronomy \& Astrophysics, University of Chicago, Chicago, IL, USA}
\email{mzzhang2014@gmail.com}

\author[0000-0002-8958-0683]{Fei Dai}
\affiliation{Department of Astronomy, University of Hawai'i, Honolulu, HI, USA}
\email{fdai@hawaii.edu} 

\author[0000-0002-6215-5425]{Qiao Xue}
\affiliation{Department of Astronomy \& Astrophysics, University of Chicago, Chicago, IL, USA}
\email{qiaox@uchicago.edu}

\begin{abstract}

\noindent Recent thermal emission observations suggest that `lava worlds'--Earth-sized exoplanets hot enough to melt their silicate surfaces--have atmospheres on a population-level.  This evidence comes from measured dayside brightness temperatures much lower than expected for a dark, maximally hot bare rock. However, these results are degenerate between a thick atmosphere advecting heat to the planet's nightside or a high albedo  reflecting incoming radiation.  This degeneracy can be resolved through visible light phase curve observations, which can constrain how much light is reflected by the planet.

Here, we present a population-level analysis of the visible-light TESS phase curves of $R\lesssim5\,R_{\oplus}$ exoplanets. We select the population by defining a `Reflection Spectroscopy Metric' meant to quantify the relative signal-to-noise of reflected light observations. We report new $>3\,\sigma$ detections of the TESS phase curve signals of LTT 9779 b and the lava worlds TOI-2431~b, TOI-561~b, 55~Cancri~e, K2-141~b, and TOI-1444~b.  We use these results and previous results from Kepler and CHEOPS to homogeneously derive geometric albedos for these planets, revealing that lava worlds generally show high albedos compared to the low geometric albedos measured for hot Jupiters at similar temperatures.  These results support an emerging picture of reflective silicate clouds on these worlds cooling their daysides. The upcoming PLATO mission is expected to tightly constrain the reflected light phase curves of lava worlds.

\end{abstract}


\section{Introduction} 

Detecting and characterizing atmospheres around terrestrial planets is a key goal for astrophysics and astrobiology.  Thermal emission observations with the James Webb Space Telescope (JWST) show that most warm ($300\,\mathrm{K}<T_{eq}<1400\,$K) rocky planets orbiting M-stars likely lack thick atmospheres \citep{coy2025population,kreidberg2025first}, perhaps barring one notable outlier \citep{August2025}. By contrast, lava worlds, Earth-sized ($R<1.9\,R_{\oplus}$) planets hot enough ($T_{eq}\gtrsim1700$\,K) to melt their dayside silicate surfaces have shown evidence for atmospheres via dayside temperatures much lower than that expected for a dark, maximally hot bare rock \citep{zieba2022k2,hu2024secondary,monaghan2025low,teske2025thick,Coy2026,Smith2026}. These low dayside temperatures suggest efficient heat transport to the nightside, molecular absorption features from gases, and/or high Bond albedo cooling their daysides, all signatures of atmospheres. Explanations of these observations that do not involve atmospheres are unlikely, as partially or fully molten silicates are expected to have low reflectivity \citep{essack2020low}, and the heat transport from magma currents is too weak to cool the dayside \citep{kite2016atmosphere}.

These low dayside temperatures complement earlier detections of the visible eclipse depths of lava worlds with Kepler that revealed dayside fluxes much \textit{larger} than expected for a dark, maximally hot bare rock (e.g., \citealt{batalha2011kepler,sanchis2013transits,zieba2022k2,singh2022probing}), suggesting additional contribution from reflected light.  An example is Kepler-10~b \citep{batalha2011kepler}, an Earth-sized ($1.5\,R_{\oplus}$, $\rho=8.8\,$g/cm$^3$) planet with a visible eclipse depth of $10.4\pm0.9\,$ppm \citep{singh2022probing}. This deep eclipse implies a very high geometric albedo of $0.74\pm0.06$ assuming negligible thermal emission in the Kepler bandpass \citep{singh2022probing}. Even a `maximally hot' surface that instantly re-radiates all incoming instellation would require a high geometric albedo of $0.56\pm0.06$.


The Transiting Exoplanet Survey Satellite (TESS) mission has provided high-quality phase curves of hot Jupiters (e.g., \citealt{wong2020tess,wong2021visible,daylan2021tess,bourrier2020optical}), placing stringent constraints on their geometric albedos, visible-band hotspot offsets, and nightside temperatures when combined with Spitzer/JWST mid-infrared observations (e.g., \citealt{arora2024constraining}). Despite its small mirror size of 10\,cm, TESS's all-sky coverage includes several targets much brighter than the typical Kepler target, and, like Kepler, TESS often observes hundreds of phase curves of a given short period planet. This large volume of data enables phase curve characterization of very small signals.

Tentative ($<3\,\sigma$) detections of the TESS phase curve signals of multiple small ($R\lesssim5\,R_{\oplus}$) planets have been reported (e.g., \citealt{crossfield2020phase,dai2021tks,valdes2022weak,patel2023cheops,murgas2022hd,dai2024earth,rubenzahl2024tess,lee2025toi}).  However, these analyses each focus on just one planet, and differ in their astrophysical assumptions and data reduction choices.  To better study the population-level constraints on small planet phase curve parameters provided by TESS, uniform analyses are needed.

Here we present a population-level analysis of the phase curves of lava worlds with TESS to better constrain their visible-NIR geometric albedos.  We present a new pipeline for analyzing TESS phase curves in Section \ref{sec:sec2} and compare results from this pipeline to published JWST data in Section \ref{sec:sec3}, where we introduce a new observability metric for reflected light observations in Section \ref{sec:RSM}.  We present new detections of TESS phase curve signals in Section \ref{sec:results} and, combining these with earlier results from the Kepler mission, discuss theoretical implications in Section \ref{sec:discussion}.  We highlight limitations of our approach and room for future work in Section~\ref{sec:futurework}, discuss future observations in Section~\ref{sec:futureobs}, and conclude in Section~\ref{sec:conclusions}.

\section{\texttt{SPARTESS}:  A New Pipeline for Analyzing TESS Phase Curves} \label{sec:sec2}

Uniform analyses are crucial to identifying potential population-level trends in low signal-to-noise observations (e.g., \citealt{baxter2020transition,mansfield2021unique,coy2025population,monaghan2026uniform}). The TESS eclipses and phase curves of small planets have been analyzed previously (e.g., \citealt{crossfield2020phase,dai2021tks,murgas2022hd,patel2023cheops,rubenzahl2024tess,dai2024earth,lee2025toi}), but lack a uniform framework needed to interpret these results.

We present \texttt{SPARTESS} (`\texttt{SPARTA} for TESS'), a new, flexible data analysis pipeline to constrain the visible light phase curves of highly-irradiated exoplanets with TESS.   The pipeline is inspired by the \texttt{SPARTA} \citep{kempton2023reflective,xue2025jwst} JWST data reduction pipeline, and builds upon several earlier analyses of TESS and Kepler phase curve data.  We start by using the PDCSAP (Pre-search Data Conditioning Simple Aperture Photometry) fluxes reported by the TESS SPOC pipeline.  Compared to the simple aperture photometry (SAP) fluxes, the PDCSAP fluxes are detrended via `cotrending basis vectors' derived from systematic effects common to all nearby stars, and are meant to preserve astrophysical signals of interest.  PDCSAP also attempts to correct for contamination/dilution due to nearby stars in the same pixel as the target.  Alternative data reduction strategies based on the full frame images and Gaia parallax data have suggested this correction may not be accurate for contaminated fields \citep{han2025hundreds}, but these data are only available at long (30 min) cadence and introduce much larger uncertainties, preventing the identification of small phase curve signals. Previous analyses of hot Jupiter TESS phase curves have suggested that PDCSAP fluxes generally produce higher-quality light curves than SAP \citep{wong2021visible}, however some specific PDCSAP sectors show anomalous noise not present in SAP (discussed more below).

For planets with outer companions, we remove the in-transit data of other planets in the system.  The phase curve signals of most companions are small enough that they are safely ignorable.

We tested linear detrending with other diagnostic data reported by the SPOC pipeline, namely the position and velocity of the center of the PSF, but found no significant correlation.

We also include an error inflation factor (unique to each sector, similar to a `jitter' term used in other studies) to ensure that we are not underestimating our uncertainties on retrieved parameters.

\subsection{Phase Curve Model}

We follow the simplified second-order sinusoidal phase curve model used in \citet{zhang2024gj} (originally derived from \citealt{cowan2008inverting}), where the planet flux at a given orbital phase $\theta$ (relative to mid-eclipse) is given by,

\begin{equation} \label{eq:phasecurve}
\begin{split}
    F_{p}(\theta) = F_{d} &+ C_{1}(\cos\theta-1) + D_{1}\sin\theta \\
                          &+ C_{2}(\cos2\theta-1) + D_{2}\sin2\theta.
\end{split}
\end{equation}
$F_{d}$ represents the planet's dayside flux/eclipse depth, whereas $C_{1}$ and $D_{1}$ control the phase curve amplitude and offset.  $\theta$ is calculated via $\theta = 2\pi\left(\left(t_{ref} - t_{0}\right)/P+0.5\right)$ where $t_0$ is the time of conjunction and P is the orbital period of the planet. $t_{ref}$ is the corrected timestamp accounting for the light travel time delay, where

\begin{equation}
    t_{ref}=t_{obs}-\frac{a}{c}(1+\cos\theta)\sin i.
\end{equation}
We calculate $\frac{a}{c}$ by fixing the stellar radius to the median value reported in the NASA Exoplanet Archive (i.e., $\frac{a}{c}=\frac{1}{c}\cdot\frac{a}{R_{\star}}\cdot R_\star$).

The eclipse and transit model lightcurves are obtained using \texttt{batman} \citep{kreidberg2015batman}, using a quadratic limb darkening law based on the methodology introduced in \citet{kipping2013efficient}.  The full light curve is calculated via
\begin{equation}
    F(\theta)=F_\star\times \left(M_{tr}(\theta)+(M_{ecl}(\theta)-1)\times F_{p} \left(\theta\right)\right),
\end{equation}
where $M_{tr}$ is the transit model, $M_{ecl}$ is the eclipse model, and $F_\star$ is a renormalization constant.

$C_{2}$ and $D_{2}$ typically capture second-order effects, such as the ellipsoidal distortion of the star and/or planet due to tidal distortion.  Doppler boosting, small changes in the star's TESS-bandpass brightness as it wobbles due to the companion, can also affect these phase curves.  For most small planets, these effects are negligible compared to the planet's signal. For example, the expected semi-amplitude of ellipsoidal variation and Doppler boosting signals for 55~Cancri~e based on equations presented in \citet{wong2021visible} are $\lesssim0.6$\,ppm and 0.02\,ppm respectively. We thus ignore these for most of our analyses.

With this framework, the (westward) phase offset can be calculated as 
\begin{equation}
    \Phi =\tan^{-1}\left(\frac{D_{1}}{C_{1}}\right)
\end{equation}
and the nightside flux as 
\begin{equation}
    F_{n}=F_{d}-2\,C_{1},
\end{equation}
with the maximum amplitude as,
\begin{equation}
    F_{max}=F_{d}-C_{1}+\sqrt{C_{1}^{2}+D_{1}^{2}},
\end{equation}

While the TESS data quality is likely not enough to diagnose potential offsets with confidence, we include $D_{1}$ in our fits and use wide uniform priors ($\mathcal{U}(-1000,1000)\,$ppm) for each phase curve parameter to allow for a diverse set of solutions and to ensure that we are actually detecting the dayside flux and phase variation.  This approach is also meant to be as data-driven as possible given the low signal-to-noise, although we discuss specific exceptions to this rule below.

This phase curve model leads to a slightly different shape than a Lambertian scatterer, which is a common approximation for modeling visible reflected light phase curves (e.g., \citealt{hu2015semi,dai2021tks,rubenzahl2024tess}).  This difference, however, is minor and likely not detectable given the signal-to-noise of TESS observations.  In addition, assuming Lambertian scattering requires knowledge of the exact proportion of reflected light to emitted light contribution and the scattering properties of the atmosphere, both of which are difficult-to-impossible to constrain observationally for small planets.

\subsection{Detrending for Stellar Activity and TESS Systematics}\label{sec:detrend}

The TESS phase curve amplitudes of small planets are small ($\lesssim$100\,ppm) compared to typical stellar variability or systematic signals, which can be  $\sim20000$\,ppm for active young stars.  The precision of phase curve parameters is thus often limited by how accurately one can remove these systematic signals.  This is even more so the case for TESS than Kepler due to the much lower photometric stability and precision.  Here, `systematics' refers to a combination of photometric variability of the host star and variations due to the telescope itself, which can often not be separated for low SNR photometry.

\subsubsection{Pre-fit Detrending Methods}

A common way to remove these systematics is by fitting a third-order spline function \citep{vanderburg2014technique}.  To preserve phase variation of the planet, the minimum break space must be longer than the orbital period of the planet.  We use the `Kepler spline' methodology introduced in \citet{vanderburg2014technique}, which iteratively fits a basis spline and tests a variety of break spaces while masking out the transit.   We use this to fit for separate splines per individual sector, as stellar signals can vary significantly on long timescales.  This approach is the most physically motivated for modeling rotation-induced stellar variation due to being a smooth and continuous function, but can miss short timescale variations brought upon either by stellar variations or TESS systematics.

Another way to remove TESS systematics is by applying a `moving median' filter to each point by taking the median of data points within half of the orbital period (e.g., \citealt{bourrier2020optical}).  The basic premise is that the median of any point on the phase curve over a full orbital period should be roughly equal (assuming no inherent variability of the planet's signal), and that any variations in the actual data should be due to stellar variability or telescope systematics.

Yet another common approach is that introduced in \citet{sanchis2013transits}, which has been used in various investigations of Kepler and TESS data \citep{zieba2022k2,singh2022probing,rubenzahl2024tess}.  This method is similar, but not equivalent, to the moving median method described above.  For each data timestamp, data within half of the orbital period is used to fit a linear regression, masking out transit events.  The detrending point is then the value of that function evaluated at the original timestamp.  We refer to this method as the `sliding linear fit' below.

\begin{figure*}
    \centering
    \includegraphics[width=0.9\linewidth]{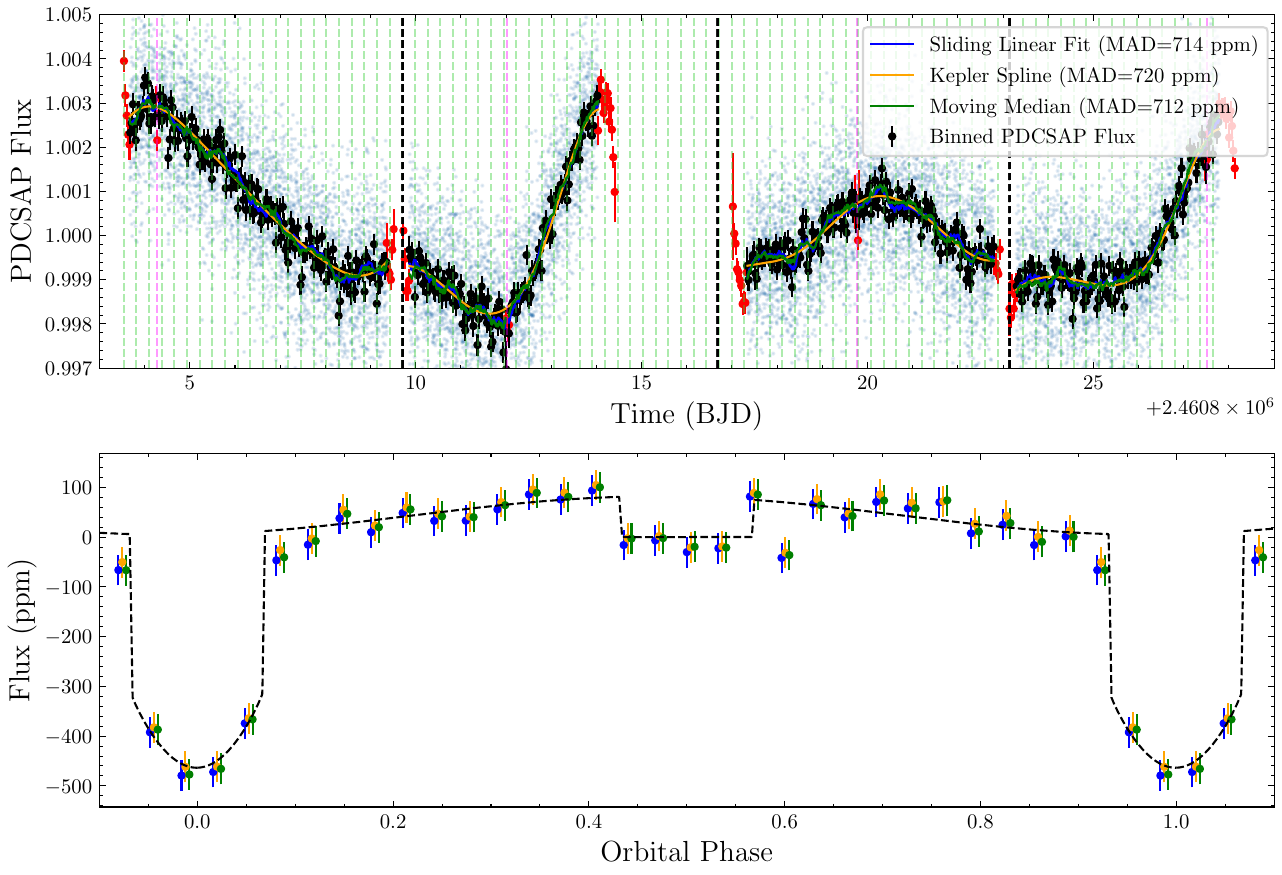}
    \caption{An example of different pre-fit stellar/systematic detrending algorithms applied to TESS Sector 92 data of K2-141.  Green dashed vertical lines indicate transits of K2-141~b with magenta indicating transits of K2-141~c, whereas black dashed lines indicate momentum dump events of TESS. Red regions indicate gap edges or transits of K2-141~c cut from our data. (bottom) The phase-folded data from all three sectors and differences using each detrending method after fitting for the full phase curve.  The different methods show very good agreement (see also Table \ref{tab:detrend_k2141}).}
    \label{fig:detrend}
\end{figure*}

In Figure \ref{fig:detrend}, we apply these detrending methods to K2-141, a young active K dwarf that hosts the lava world K2-141~b and a sub-Neptune K2-141~c with a grazing transit.  After fitting a full phase curve model, the spline function shows the largest median absolute deviation (MAD) of the residuals of 720 ppm, compared to the nearly-identical MAD for the sliding linear fit and moving median methods (714 and 712 ppm).  While the spline function is the most physically motivated, it can miss small oscillations in systematics on the order of $\sim1$\,d, which we see evidence for in the full light curve.  

The key disadvantage to these pre-fit detrending algorithms is that they require precise knowledge of the transit ephemerides, as transit events must be masked in the spline and sliding linear fit methods. They also cannot capture the inherent uncertainty of the detrending methods themselves. As noted in \citet{singh2022probing}, these pre-fit detrending methods also tend to fail when the timescale of variability is on the order of the planet's orbital period, typically for young active stars hosting planets with orbital periods $\gtrsim$1 day.

\begin{deluxetable*}{lcccccc}
\tabletypesize{\scriptsize}
\tablenum{1}
\tablecaption{TESS Phase Curve Parameters Derived for K2-141~b Using Different Stellar Variability/Systematics Detrending Methods}\label{tab:detrend_k2141}
\tablewidth{0pt}
\tablehead{
\colhead{Method} & \colhead{Dayside Flux} & \colhead{Phase Offset} & \colhead{Nightside Flux} & \colhead{Radius} & \colhead{$q_{1}$} & \colhead{$q_{2}$}\\
\colhead{} & \colhead{(ppm)} & \colhead{(degrees east)} & \colhead{(ppm)} & \colhead{($R_{p}/R_{\star}$)} & \colhead{} & \colhead{}}
\startdata
Sliding Linear Fit & $82.1^{+19.8}_{-19.4}$ & $14^{+13}_{-11}$ & $3.1^{+19.5}_{-19.2}$ & $0.0200^{+0.0005}_{-0.0006}$ & $0.22^{+0.21}_{-0.12}$ & $0.48^{+0.33}_{-0.32}$ \\
Kepler Spline & $81.4^{+19.6}_{-19.8}$ & $16^{+12}_{-11}$ & $2.6^{+19.3}_{-19.7}$ & $0.0199^{+0.0005}_{-0.0006}$ & $0.23^{+0.23}_{-0.12}$ & $0.49^{+0.34}_{-0.31}$ \\
Moving Median &$84.3^{+19.8}_{-19.7}$ & $16^{+12}_{-11}$ & $10.4^{+20.2}_{-20.0}$ & $0.0200^{+0.0005}_{-0.0006}$ & $0.22^{+0.22}_{-0.11}$ & $0.48^{+0.33}_{-0.31}$  \\
GP (M\'atern 3/2 Kernel) &  $75.7^{+19.2}_{-20.0}$ & $16^{+12}_{-11}$ & $2.0^{+19.8}_{-20.3}$ &  $0.0200^{+0.0005}_{-0.0006}$ & $0.23^{+0.23}_{-0.12}$ & $0.49^{+0.33}_{-0.31}$ \\
GP (SHO Kernel) &  $79.7^{+19.9}_{-19.8}$ & $14^{+12}_{-11}$ & $-1.6^{+19.9}_{-19.8}$ & $0.0199^{+0.0005}_{-0.0006}$ & $0.22^{+0.21}_{-0.12}$ & $0.51\pm0.32$  \\
GP (Rotation Kernel) & $80.8^{+19.7}_{-19.6}$  & $14^{+12}_{-11}$ & $0.0^{+19.1}_{-19.6}$ & $0.0200^{+0.0005}_{-0.0006}$ & $0.22^{+0.23}_{-0.12}$ & $0.49^{+0.32}_{-0.30}$   \\
\enddata
\tablecomments{In each case the dayside flux is detected at $>3\,\sigma$. $q_{1}$/$q_{2}$ refer to limb-darkening coefficients based on the parameterization of \citet{kipping2013efficient}. Note that phase offset is derived from the $D_{1}$ parameter (Equation \ref{eq:phasecurve}).}
\end{deluxetable*}

\subsubsection{Gaussian Processes}

Gaussian processes (GPs) are commonly used to fit systematic trends in light curve data and offer a more flexible realization of quasi-periodic variations due to the star or systematics than the pre-fit methods described above.   They can also capture uncertainty in the detrending process itself, and often increase the uncertainty in retrieved parameters to account for correlated/systematic noise. However, one must use caution when comparing best-fit results using GPs to the pre-fit detrending methods above.  Since GPs are inherently more flexible to the exact properties of systematics/correlated noise (and are effectively fitting to this exact noise), commonly used metrics like the median absolute deviation or RMS of residuals for the best-fit light curves will typically be better for GP noise models. 

We use three commonly-used kernels implemented in \texttt{celerite2} \citep{celerite2} to address correlated noise in TESS data.  The first is the M\'atern 3/2 kernel, represented as,
\begin{equation}
    k(\tau) = \sigma^2 \left( 1 + \frac{\sqrt3\tau}{\rho} \right)\exp\left(-\frac{\sqrt3\tau}{\rho} \right),
\end{equation}
which assumes the noise (amplitude $\sigma$) over a given time-lag $\tau$ decays exponentially with timescale $\rho$.  This model is generally the most flexible for mitigating general correlated noise but is the least physically-based and has two free parameters per kernel.

The second is the Simple Harmonic Oscillator (SHO) kernel in the form of
\begin{equation}
    S(\omega) = \sqrt{\frac{2}{\pi}} \frac{S_0\,\omega_0^4}
{(\omega^2-{\omega_0}^2)^2 + {\omega_0}^2\,\omega^2/Q^2}.
\end{equation}
The variables $\omega_0$, $S_0$, and $Q$ are related to the undamped period of the oscillator $\rho$, amplitude $\sigma$, and a new variable $\tau$ representing the damping timescale, where $\omega_0=2\pi/\rho$, $Q=\omega_0\times\tau/2$, and $S_0=\sigma^2/(\omega_0Q)$. This kernel is generally more physically representative of spot-induced rotational modulation than the M\'atern kernel and has three free parameters per kernel.

The third is the rotation kernel, which is a mixture of two SHO terms at periods $P$ and $P/2$, designed to model rotational modulation of the star.  This kernel adds a secondary quality factor $Q_{0}$, a variable for the difference between quality factors of the first and second modes $dQ$, and a fractional amplitude of the secondary mode $f$, giving five free parameters per kernel.

We fit for separate GP hyperparameters for each sector.  The magnitude and timescale of stellar variability can change drastically over sectors separated by a large gap in time.  Even for adjacent sectors, the star does not lie on the same pixel of the detector and has a slightly different background, meaning that the systematics due to the telescope can vary even if stellar variability is similar.  Fitting for separate hyperparameters per sector generally gave more precise retrieved phase curve parameters than a global GP fit despite introducing more free parameters, indicating that the per-sector fit is a better model of the underlying systematics.  

Fits to TESS data of K2-141 using GPs generally show good agreement with the pre-fit detrending methods (Table \ref{tab:detrend_k2141}). 
We restrict ourselves to using GP detrending models only in the case of strong correlated noise or rotational modulation.  We quantify how time-correlated systematics affect retrieved phase parameters with injection-recovery tests in Appendix \ref{ap:injrec}.  

\subsection{Gap Trimming}

TESS experiences `momentum dump' events every 2--6~days that are accompanied by sharp changes in systematics.  The PDCSAP fluxes automatically remove parts of the lightcurve most affected by these dumps, but there are often residual ramps before and after these dumps and the end/beginning of sectors due to the pointing and thermal settling of the telescope. 

Data near large gaps (often close in time to momentum dumps) sometimes show heightened levels of systematic ramping that are not well-handled by simple detrending algorithms. Prior to Sector 56, the dumps occurred during observations with no gap in data collection, whereas following Sector 56 they are generally separated by $\gtrsim0.2$ day gaps.  While the PDCSAP fluxes generally seem to remove most systematic trends caused by such events, residual systematic trends persist in some of these during-observation momentum dumps. To deal with these issues, we use an adaptive algorithm to identify data most affected by residual ramps and discontinuities near large gaps in data and momentum dump events. We use this instead of the flat 0.5 day gap trimming done in similar studies \citep{wong2020systematic} to preserve as much of the data as possible while removing statistically anomalous regions of noise.

\paragraph{Adaptive Clipping}
In removing data near large gaps, we use a floor of $P/2$ (where $P$ is the orbital period of the planet).  This is because our detrending algorithms use a minimum baseline of $\pm P/2$ when fitting; data within $P/2$ of large gaps do not have the full $P$ baseline needed for our detrending algorithms and so are cut. 

Beyond this floor, we test whether the data next to each gap behave like the rest of the sector, and cut further regions only where they do not. Working outward from the floor in windows of 0.25\,d, on 10-minute bins with all known transits masked, we measure four properties of each window:
\begin{enumerate}
    \item its mean flux relative to the following day of data (meant to detect a sudden jump/discontinuity in flux or the tail of a settling ramp),
    \item its linear slope (to determine whether there is a ramp),
    \item the scatter of its residuals after detrending with the sliding linear fit, relative to the interior regions of the sector (meant to detect structure shorter than $P/2$, which the detrending cannot remove), and
    \item its median sky background (\texttt{SAP\_BKG}) relative to the interior (scattered light, independent of the flux).
\end{enumerate}
The offset and slope are judged against the same quantities measured in 300 randomly placed windows in the interior regions of the same sector, so that the thresholds scale with each star's own variability: a slope of 1000\,ppm\,d$^{-1}$ is anomalous on a quiet star like TOI-2431 but not on one with strong rotational modulation like K2-141. A window is marked as anomalous if its offset or slope exceeds 3 times the interior scatter of that quantity, if its residual scatter exceeds 1.3 times the interior value, or if its background exceeds twice the interior median. Each anomaly extends the cut by 0.1\,d, and the next window is tested, up to a maximum of 2\,d. For sectors with strong stellar variability, where these clipping thresholds become wide, we use a stricter variant that extends the cut when a window reaches a lesser `warning' level (2 times the interior scatter, 1.2 times the residual scatter, or 1.5 times the background).

\paragraph{Momentum Dumps:}
Momentum dump events are identified from the SPOC data quality flags. For each dump that occurs during observations, we measure the flux jump/offset across it as the difference between lines fitted to the 0.25\,d before and after the dump. Both are evaluated at the dump time, so that a smooth stellar slope is not mistaken for a jump. The data on either side are then scanned iteratively outward with the same four tests as above, using a detrending that is not allowed to cross the dump. Where the step exceeds 3 times its interior scatter, the detrending window is split at the dump rather than the data being excluded, since excluding data cannot repair a discontinuity. Dumps that fall inside a data gap are handled by the gap-edge procedure.  The lightcurves for each sector are visually inspected for remaining correlated noise, both near momentum dumps and gaps and also the interior regions of the lightcurve.

To remove remaining outliers, we performed sigma clipping on the phase-folded data by applying a median filter with a width of 0.005 orbital phase, ignoring the ingresses and egresses. This approach deals with remaining exponential ramps, flaring events, and/or cosmic ray hits that may be missed from our detrending and gap/dump-trimming algorithms.  We tested using asymmetric bounds for sigma clipping (i.e. removing points 3$\,\sigma$ above and 5$\,\sigma$ below the median as done in \citealt{zieba2022k2}) but found that symmetric clipping of points $\pm4\,\sigma$ from the median generally gave the best quality light curves.

\section{Comparison to Previous Results} \label{sec:sec3}

Here, we compare results from our TESS analyses to published JWST NIRISS phase curve observations.

\subsection{LTT~9779~b}

LTT~9779~b \citep{jenkins2020ultrahot} is an ultra-hot Neptune-sized planet ($T_{irr}=2800\,$K\footnote{The irradiation temperature $T_{irr}$ is the sub-stellar temperature assuming zero albedo and zero heat redistribution, related to equilibrium temperature as $T_{irr}=\sqrt{2}\,T_{eq}$.}, $R=4.7\,R_{\oplus}$, $M=29\,M_{\oplus}$). Secondary eclipse observations with CHEOPS \citep{saha2025high} and phase curve observations with JWST NIRISS \citep{coulombe2025highly}, combined with constraints on the planet's thermal emission from Spitzer IRAC \citep{dragomir2020spitzer}, have revealed an extremely high geometric albedo ($A_{g,CHEOPS}=0.73\pm0.11$, $A_{g,NIRISS}~=~0.50~\pm~0.07$) and moderate Bond albedo ($A_{B}~=~0.31~\pm~0.06$).  This high albedo is theorized to be due to reflective silicate (Mg$_2$SiO$_4$ or MgSiO$_3$) clouds. \citet{crossfield2020phase} analyzed the Sector 2 TESS phase curve of LTT~9779~b, recovering an eclipse depth of $55_{-21}^{+24}$~ppm ($2.6\,\sigma$ from zero), roughly in agreement with the NIRISS shortwave (0.6--1.0\,$\mu$m) data ($\sim80$\,ppm), which covers the same wavelength range as TESS (0.6--1.0\,$\mu$m).  However, they were unable to detect a significant phase variation or offset, retrieving a high nightside flux of $33^{+24}_{-20}$~ppm.  Here, we use three new sectors of TESS data (29, 69, and 96) to refine LTT~9779~b's phase curve parameters.  We compare our results to the NIRISS shortwave (0.6--1.0$\,\mu$m) reduced data presented in \citet{coulombe2025highly}.  While these data cover the same wavelength range as TESS, the NIRISS data has slightly more contribution from the longest $\sim$0.9--1.0$\,\mu$m wavelengths due to the difference in throughput between NIRISS and TESS, leading to a slightly higher bandpass-integrated eclipse depth.

\begin{figure*}
    \centering
    \includegraphics[width=1.0\linewidth]{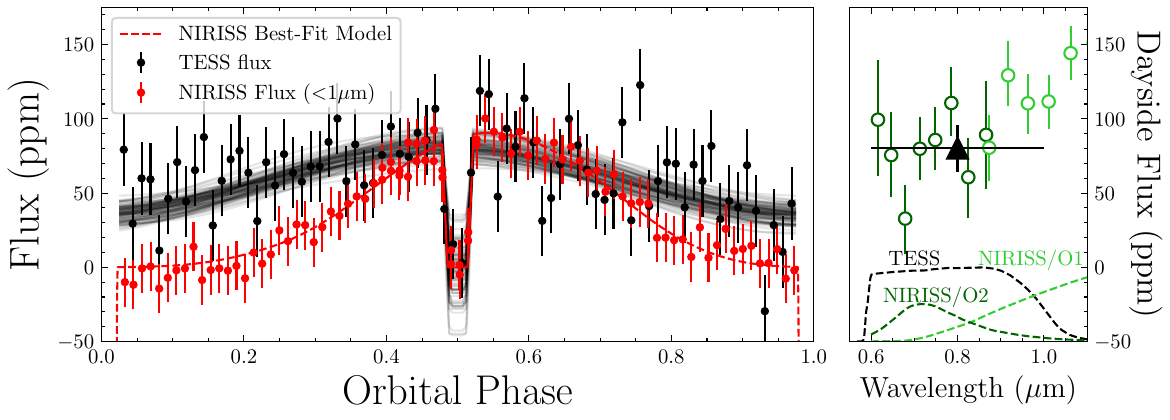}
    \caption{(left) The sectors 2, 29, 69, and 96 TESS phase curve of the ultrahot Neptune LTT~9779~b using \texttt{SPARTESS} compared to the shortwave (0.6--1.0$\,\mu$m) JWST/NIRISS white light phase curve presented in \citet{coulombe2025highly}.  While the eclipse depths show good agreement, we do not recover the near-zero nightside flux and strong westward offset observed with NIRISS.  We compare the retrieved TESS dayside flux (black) to the JWST/NIRISS shortwave dayside emission spectrum (green) in (right).  The throughput functions of each instrument are shown as dashed lines.}
    \label{fig:LTT9779}
\end{figure*}

Our gap trimming algorithm identified four regions of heightened noise cut from our analysis:  [[2458378.5310, 2458378.6310], [2459098.1945, 2459098.3945], [2459108.3904, 2459108.4904], [2459112.1637, 2459112.2637]] BJD. Using sliding linear fit detrending, we retrieve a dayside flux of $80.2\pm15.9$\,ppm, in good agreement with the dayside spectrum retrieved for NIRISS data. The host star is relatively inactive with a long rotation period of $\sim45\,$d \citep{jenkins2020ultrahot}, and we saw no evidence for significant time-correlated noise after applying this detrending algorithm.  The resulting phase curve and comparison to the NIRISS shortwave data is shown in Figure \ref{fig:LTT9779}.  We are unable to recover the significant westward offset reported in \citet{coulombe2025highly}, instead preferring a negligible offset of $5\pm10$$\degree$ east. We also retrieve a nightside flux of $35.3^{+16.0}_{-15.6}$\,ppm, which is $2.3\,\sigma$ consistent with zero but noticeably higher than the negligible nightside flux observed with NIRISS.  The nightside flux is also physically implausible, as even maximal heat redistribution to the nightside (with 0 Bond albedo) would only result in a flux of $\sim7\,$ppm.  These results highlight that out of the three major phase curve parameters: dayside flux/secondary eclipse depth, offset, and nightside flux, TESS data is likely most useful in constraining the dayside flux due to biases introduced by the aggressive detrending required.  We found that this high nightside flux persists regardless of detrending method used (including GPs) and whether the SAP or PDCSAP fluxes were used.  A leave-one-out analysis results in similar nightside fluxes of $35\pm19\,$\,ppm, $43\pm18\,$ppm, $32^{+18}_{-17}$\,ppm,  and $30\pm18\,$ppm when excluding data from sector 2, 29, 69, and 96, respectively.  Furthermore, enforcing zero nightside flux while using GP detrending leads to a low dayside flux ($\sim48\,$ppm) inconsistent with the NIRISS dayside spectrum, indicating the non-zero nightside flux is likely a real feature of the given data quality.

\subsection{WASP-121~b}

WASP-121~b \citep{delrez2016wasp} is an ultra-hot Jupiter ($T_{irr}=3400\,$K, $R=19.5\,R_{\oplus}$, $P=1.27\,$d) also observed over a full phase curve with JWST NIRISS/SOSS \citep{splinter2025precise}.  The Sector 7 TESS phase curve was analyzed independently by three teams, retrieving 2$\,\sigma$ consistent dayside fluxes of $482^{+39}_{-41}$\,ppm \citep{daylan2021tess}, $486\pm59$\,ppm \citep{wong2020systematic}, and $419^{+47}_{-42}$\,ppm \citep{bourrier2020optical}.  \citet{daylan2021tess} also report a nightside flux of $65^{+40}_{-37}\,$ppm.  \citet{daylan2021tess} and \citet{wong2020systematic} used the raw SAP fluxes in their analysis whereas \citet{bourrier2020optical} used PDCSAP.  Here, we take advantage of five new sectors of TESS data (33, 34, 61, 87, 88) and compare our results to the NIRISS shortwave ($<1.1\,\mu$m) data presented in \citet{splinter2025precise}.  For these fits, we also include the $C_{2}$ parameter (Equation \ref{eq:phasecurve}) in our phase curve model, as WASP-121~b is massive enough to induce moderate ($\sim18\,$ppm, \citealt{wong2020systematic}) ellipsoidal variation on the host star.  The planet should also induce a small $\sim1.8\,$ppm Doppler boosting effect on the phase curve.

We initially used the moving median detrending algorithm to remove long-term stellar and instrument variations.  This resulted in a dayside flux of $424^{+16}_{-17}$ ppm and a (non-negligible) nightside flux of $67\pm17$ ppm.  However, we found high amounts of time-correlated noise in the resulting lightcurve, likely because the stellar rotation period is close to the orbital period of the planet ($P_{rot}=1.13$\,d, $P_{b}=1.27$\,d, \citealt{bourrier2020hot}). Our moving median detrending algorithm can inherently not capture this sub-orbital-period variability due to the star.  We also noticed that sector 7 showed much stronger correlated noise in the PDCSAP data than SAP, and thus we use SAP for sector 7 only (corrected with a crowding factor of 0.9113, trimmed to the same timespans as PDCSAP) in the fits below.

To address the correlated noise, we use a GP model with \texttt{celerite2}'s M\'atern-3/2 kernel using wide uninformative priors.  The GP model inflates errors on our retrieved parameters, with a dayside flux of $410\pm18\,$ppm (a 22$\,\sigma$ detection) and a nightside flux of $32^{+36}_{-37}\,$ppm.  The new nightside flux is now consistent with the expected TESS flux based on the reported nightside effective temperature retrieved from the NIRISS phase curve ($\sim2$\,ppm  from $T_{night}=1562\,$K, \citealt{splinter2025precise}).
We retrieve a negligible phase offset of $2.1\pm5.1$$\degree$ east (after accounting for a Doppler boosting signal of 1.8\,ppm, \citealt{wong2020systematic}), consistent with zero and $\sim2\times$ more precise than the offset of $10.5\pm9.9\degree$ reported for NIRISS Order 2 \citep{splinter2025precise}.  We also retrieve an ellipsoidal distortion signal ($C_{2}$) of $20\pm10\,$ppm, consistent with the theoretical expectation of 18\,ppm.

\section{Reflection Spectroscopy Metric} \label{sec:RSM}

To aid future phase curve or eclipse observations of small planets in the visible, we define a priority metric that estimates -- for a given albedo and observing time -- the expected signal-to-noise of reflected light observations.  Here, we present a definition for the reflection spectroscopy metric (RSM), building on the transmission and emission spectroscopy metrics introduced in \citet{kempton2018framework}.  We define RSM as
\begin{equation}\label{eq:RSM}
    \mathrm{RSM}\propto A_{g}\left(\frac{R_{p}}{a}\right)^2 10^{-m_{Gaia}/5},
\end{equation}
where $A_{g}$ is the assumed geometric albedo in a given bandpass, $R_{p}$ is the planet radius, $a$ is the planet's semi-major axis, and $m_{Gaia}$ is the host star's Gaia magnitude.  Of the 5 host star magnitudes available by default in the NASA Exoplanet Archive (Gaia, Johnson \textbf{V}, 2MASS \textbf{J}, \textbf{H}, and \textbf{K}) Gaia most accurately captures the relative broadband brightness of stars in the visible-NIR most amenable to reflected light observations. Thus we use the $10^{-m_{Gaia}/5}$ factor to represent the relative photon noise. $\left(R_{p}/{a}\right)^2$ is the maximum planet-to-star reflected light contrast ratio assuming $A_{g}=1$, and can also be expressed as $\left(R_{p}/R_{\star}\times (a/R_{\star})^{-1}\right)^2$.  This parameterization assumes an equal observing time per target, and that every photon is used to measure the reflected light signal (i.e. photons are equally useful regardless of orbital phase), which is typically the case in phase curve observations.  The largest uncertainty in this parameterization is $A_{g}$, which cannot be known a priori. $A_{g}$ also has a wavelength-dependence (e.g, \citealt{marley1999reflected}), controlled by atmosphere and cloud composition. For this work, we make simplifying assumptions for $A_{g}$, based on previous observational and/or model results. Specifically, we assume targets in the hot-to-ultra-hot regime ($T_{irr}>2400\,\mathrm{K}$) generally have high albedos (0.5 for super-Earths, 0.7 for sub-Neptune to Neptunes) but low albedos (0.1) for cooler planets, based on the low albedos inferred for the typical Kepler rocky/Neptune-sized planet \citep{demory2014albedos,sheets2017average,jansen2018kepler}. As an alternative, we also include an RSM$^*$ that assumes equal albedo. We normalize RSM/RSM$^{*}$ to the signal-to-noise of the dayside flux measured for LTT 9779 b over four TESS Sectors, so that LTT 9779 b has an RSM/RSM$^{*}$ of 5.05.  Here, we focus on the reflected signals of small planets, which are expected to dominate over their thermal signals at  wavelengths $<1\,\mu$m.  (Ultra-hot Jupiters, such as WASP-121~b, can have significant thermal emission signals in the TESS bandpass that complicate the interpretation of RSM.)  We compare the scaling predicted by RSM to Kepler observations of lava worlds in Appendix Section~\ref{ap:RSM_comparison}.

In this work, we investigate the TESS phase curves of the top 25 RSM and RSM$^{*}$ targets.  We also investigate targets where tentative ($>2\,\sigma$) phase curve signals have been reported in TESS or Kepler: TOI-1347~b \citep{rubenzahl2024tess}, Kepler-407~b, Kepler-10~b, Kepler-78~b, K2-106~b, K2-229~b, and K2-312~b \citep{singh2022probing}.  A full list of the 37 targets analyzed in this work is shown in Table \ref{tab:RSM}.  A few targets (mostly hot Neptunes around dimmer stars) have only long cadence (30 min or 10 min) data available for certain sectors.  For these targets, we included the SPOC pipeline data from long-cadence sectors and super-sampled the data to a resolution of 2 minutes. 

Our workflow was as follows: we ran initial \texttt{dynesty} fits with a small number of live points (200) using the moving median detrending method using the NASA Exoplanet Archive ephemerides. This method is less sensitive to ephemeris precision than the sliding linear fit or spline methods, which require masking transits. If the target indicated a potential phase curve signal, we followed up with visual and statistical inspection of the light curves and a larger number of live points (1000, see below) using dynamic nested sampling. 
Notably, almost all phase curves analyzed showed a dayside flux above zero, but most lack the required precision for confident ($>\,3\sigma$) detections of the dayside flux.

TOI-1444~b is the lowest RSM (0.42) target where we confidently ($>3\sigma$) detected a dayside flux in TESS.  This is due to its high apparent geometric albedo ($\sim0.6$) compared to values assumed in Table \ref{tab:RSM} and large number of TESS sectors (24).  Accounting for the potential diversity in geometric albedo, we recommend an RSM of 0.5 as a lower limit for investigating visible light phase curves with TESS-like telescopes.  Future missions may allow for lower RSM targets: see Section \ref{sec:futureobs}.  Several Kepler and K2 targets with confident eclipse detections have low RSM values due to Kepler's high photometric precision.  For example, \citet{singh2022probing} report $>3\,\sigma$ detections of the dayside fluxes of Kepler-10~b (RSM: 0.19), Kepler-78~b (0.33), K2-141~b (1.11), K2-106~b (0.23), K2-131~b (0.52), and Kepler-407~b (0.07).

\begin{deluxetable*}{lcccccccc}
\tablenum{3}
\tablecaption{Top 25 Reflection Spectroscopy Metrics for Small Planets ($R<6\,R_{\oplus}$)}\label{tab:RSM}
\tablewidth{0pt}
\tablehead{\colhead{Planet} & \colhead{$R_{p}$} & \colhead{$T_{irr}$} & \colhead{Period} & \colhead{$(R_{p}/a)^{2}$ (ppm)} & \colhead{$m_{gaia}$} & \colhead{RSM} & \colhead{RSM$^{*}$} & \colhead{Num. TESS Sectors}\\
 & \colhead{$(R_{\oplus})$} & (K) & (d) & \colhead{[Max. Ref. Signal]} & \colhead{} & & \colhead{(Equal Albedo)} & \colhead{}}
\startdata
LTT~9779~b$^{\dagger}$ & 4.72 & 2800 & 0.79 & 143 & 9.60  & 5.05 & 5.05 & 4\\
55~Cancri~e$^{\dagger}$ & 1.86 & 2770 & 0.74 & 24 & 5.73 & 3.65 & 5.11 & 4 \\
TOI-2431~b$^{\dagger}$ & 1.53 & 2920 & 0.22 & 108 & 10.32  & 1.94 & 2.72 & 5\\
TOI-849~b & 3.68 & 2810 & 0.77 & 102 & 12.06  & 1.16 & 1.16 & 5 \\
K2-141~b$^{\dagger}$ & 1.51 &  2970 & 0.28 & 74 & 10.72 & 1.11 & 1.56 & 3 \\
TOI-7008~b$^{1}$ & 4.95 & 2710 & 0.84 & 177 & 13.45 & 1.06 &1.06 & 5 \\
TOI-2260~b & 1.62 & 3690 & 0.35 & 51 & 10.25 & 0.95 & 1.32 & 6\\
TOI-431~b & 1.28 & 2630 & 0.49 & 23 & 8.79 & 0.85 & 1.19 & 4 \\
TOI-332~b$^{*}$ & 3.20 & 2650 & 0.78 & 73 & 12.06 & 0.83 & 0.83 & 8 \\
HD~213885~b & 1.75 & 3000 & 1.01 & 14 & 7.80 & 0.79 & 1.10 & 8 \\
TOI-1807~b & 1.50 & 2400 & 0.55  & 27 & 9.68 & 0.66 & 0.93 & 4 \\
HD~20329~b & 1.72 & 3020 & 0.93 & 17 & 8.59 & 0.66 & 0.93 & 4\\
TOI-561~b$^{\dagger}$ & 1.40 & 3260 & 0.45 & 31 & 10.01 & 0.65 & 0.91 & 5\\
TOI-3261~b & 3.82 & 2430 & 0.88 & 90 & 13.06 &0.65 & 0.65 & 11\\
K2-100~b & 3.57 & 2660 & 1.67 & 26 & 10.41 & 0.62 &  0.62 & 4 \\
HD~3167~b & 1.60 & 2500 & 0.96 &15 & 8.76 &0.60 & 0.84 & 1\\
TOI-2196~b & 3.51 & 2630 & 1.19 & 45 &11.83 & 0.57 & 0.57 & 4\\
K2-131~b & 1.69 & 3140 & 0.37 & 59 & 11.90  & 0.52 & 0.72 & 4 \\
HD 80653 b (K2-312~b) & 1.61 & 3480 & 0.72 & 17 & 9.31 & 0.49 &    0.69 & 2     \\
TOI-1836~c & 2.60 & 2840 & 1.77 & 13  & 9.64 & 0.44 & 0.44 & 13\\
TOI-1798.02 & 1.46 & 2990 & 0.44 & 33 &11.07 & 0.43 & 0.60 & 7\\
TOI-1444~b$^{\dagger}$ &1.42 & 3010 & 0.47 & 27 & 10.64 & 0.42 &0.59 & 24\\ 
K2-390~b & 4.66 & 2370 & 3.31 & 64  & 13.34 &0.40 &0.40 & 3\\
TOI-1839~b & 2.22 & 2270 & 1.42 & 19 & 10.73 &0.40 &0.40 & 3\\
NGTS-4 b & 3.19 &2330 & 3.19 & 51 & 12.91 & 0.39 &0.39 & 4\\
\hline\hline
TOI-6255 b & 1.08 & 1940 & 0.24 & 74 & 11.63 & 0.15 & 1.03 & 3 \\
TOI-3862~b & 5.53 & 2180 & 1.56 & 86 &12.23 & 0.13 & 0.90 & 4\\
TOI-1410~b & 3.10 & 1920 & 1.22 & 41 & 10.87 & 0.11 & 0.80 & 3\\
GJ 367 b &  0.70 & 1930 &  0.32   & 18 & 9.15 & 0.11 & 0.76  & 7  \\
TOI-1288~b & 4.97 & 1700 & 2.70 & 32 &10.45 & 0.11  & 0.76 & 13\\
HAT-P-11 b & 4.90 & 1250 & 4.89 & 16 & 9.15 & 0.10 & 0.68 & 11 \\
\hline\hline
Kepler-78~b & 1.20 & 3140 & 0.36 & 32 & 11.53 & 0.33 &0.47 & 9\\
TOI-1347~b & 1.81 & 2400 & 0.85 & 20 & 11.21 & 0.25 & 0.34 & 39 \\
K2-229b & 1.26 & 2720 & 0.58 & 17 & 10.79 & 0.24 &0.34 & 2\\
EPIC 220674823 b (K2-106 b) & 1.61 & 3260 & 0.57 & 27 & 11.95 & 0.23 & 0.32 & 2\\
Kepler-10 b & 1.47 & 3090 & 0.84 & 14 & 10.92 & 0.19 & 0.27 & 11 \\
Kepler-407 b & 1.19 & 3150 & 0.67 & 13 & 12.51 & 0.07 & 0.10 & 11 \\
\enddata
\tablecomments{The top 25 reflection spectroscopy metrics for small planets with irradiation temperatures $>1000$\,K. The middle includes targets in the top 25 RSM$^*$ values (RSM assuming an equal albedo for every target) but not in the top 25 RSM values.  The bottom panel includes planets with previous $>2\,\sigma$ detections in Kepler or TESS also investigated in this study. Data are taken from the NASA Exoplanet Archive (queried Jul 20, 2026). $^{\dagger}$ indicates planets where we detect a dayside flux at $>3\,\sigma$, where $^{*}$ indicates tentative (2--3$\,\sigma$) detections.  As well as having the highest geometric albedo confidently measured for an exoplanet, LTT~9779~b represents the highest expected signal-to-noise for reflected light observations due to its large planet-to-star radius ratio, close-in orbit, and bright host star.
$^{1}$ The detection of this planet was made with only TESS data \citep{lafarga2025} and thus may still be considered a planet candidate.}
\end{deluxetable*}

A different `Reflection Spectroscopy Metric' was defined recently in \citet{winterhalder2026spinning} focusing on ground-based high-resolution cross-correlation observations of hot Jupiters.  However, this definition differs from the transit photometry/spectroscopy-focused definitions of ESM and TSM presented in \citet{kempton2018framework}, which are adopted here.

\section{Results} \label{sec:results}
\subsection{$\geq3\,\sigma$ Detections}
Here, we report $\geq3\,\sigma$ detections of the TESS dayside flux signals of small planets when fixing the nightside flux to zero.  We consider this a physically motivated prior as the maximum nightside signals for these planets are small ($\lesssim6$\,ppm), and mid-IR phase curve observations imply small $2\,\sigma$ upper limits [0.8\,ppm for K2-141~b \citep{zieba2022k2}, 0.04\,ppm for TOI-561~b \citep{boucher2026a}, and 0.3\,ppm for 55 Cancri e \citep{mercier2022revisiting}]. A high dayside flux also inherently implies a small nightside flux; either due to poor heat redistribution (a hot dayside) or a high albedo (a reflective dayside) cooling the dayside and nightside.  However, we adopt the dayside fluxes when allowing for free nightside flux (resulting in lower precision) as our main results as a conservative estimate on the true dayside flux, which range from $2.5\,\sigma$ to $4.0\,\sigma$ from zero.  We find that for each of these fits that the retrieved nightside flux is 1$\,\sigma$ consistent with zero.

\subsubsection{Additional Validation Steps}

For planets with short eclipses relative to their full orbital periods, the retrieved dayside flux can be largely influenced by the out-of-eclipse shape of the phase curve, rather than the eclipse itself.  For example, \citet{lee2025toi} find tentative evidence for a dayside flux detection for TOI-6324~b ($42\pm28\,$ppm) despite the in-eclipse data being \textit{higher} than the out-of-eclipse data.  This shape may arise from uncertain detrending of stellar variability, and may lead to false positive detections of the dayside flux. As an extra validation step, we focus on the eclipse only to ensure that our results are not biased from the shape of the phase curve.  For these `eclipse-only' fits we generally fix orbital parameters (as they cannot be uniquely constrained) whereas for the full phase curve fits we use Gaussian priors informed by previous studies. The priors and full posteriors of our fits are shown in Appendix Table \ref{tab:lttwasp}. We require that: 1.~the eclipse is detected at greater than 2$\,\sigma$ confidence and 2.~the eclipse depth is 1$\,\sigma$ consistent with the retrieved dayside flux from the full phase curve. This method is expected to slightly underestimate the dayside flux compared to retrievals using the full phase curve: the eclipse alone cannot constrain the phase variation, and the dayside flux inside the eclipse is slightly higher than the eclipse depth itself.

In addition, for planets with short eclipses relative to their orbital period, random systematics may line up to create a false positive eclipse-like event.  If the overall phase curve shape is flat, this would be accompanied by a similarly high, likely nonphysical, nightside flux.  This problem gets worse when the eclipse is brief relative to the orbit period. 

An example is HD 20329 b, a moderate RSM (0.66), $R=1.7\,R_{\oplus}, M=7.4\,M_{\oplus}, T_{irr}=3030\,\mathrm{K}, P=0.93$\,d \citep{murgas2022hd} lava world recently observed in secondary eclipse with JWST MIRI LRS (M. Weiner Mansfield et al., in preparation). The TESS Sectors 42+43 phase curve was previously analyzed in \citet{murgas2022hd}, revealing a deep eclipse ($20\pm14$\,ppm) detected at low confidence.  This value is notably higher than the planet's maximum reflected signal of 16.6 ppm.  Here we include new data from sectors 70 and 71.  We masked regions over [2459490,~2459492] and [2460221,~2460223] BJD that showed high amounts of anomalous time-correlated noise. We retrieve a high dayside flux of $33\pm9\,$ppm ($3.8\,\sigma$ above zero) and a high nightside flux of $22\pm8\,$ppm.  These high fluxes persist regardless of detrending algorithm, including using Gaussian process models, and are consistent with eclipse-only fits. The dayside flux is unphysically high, requiring a geometric albedo of $\sim2\pm0.5$ or a temperature $1.3\pm0.1\times$ the theoretical maximum. The nightside flux is also unphysically high, requiring a temperature $\sim1.5\pm0.2\times$ that of a planet with zero Bond albedo and full heat redistribution. 

We thus discard fits with anomalously high nightside fluxes.  This does not mean that these signals are not real, rather that the data are currently too anomalous and noisy to claim detection.

\subsubsection{TOI-2431~b}

TOI-2431~b \citep{tacs2026earth} is a recently-validated ultra-short period lava world ($T_{irr}=2900\,$K, $P=0.22\,$d).  The planet's large planet-to-star radius ratio, host star brightness, and short orbital period make it a prime observing target, with the largest maximum reflected light signal size (110 ppm) of any lava world ($R<1.9\,R_{\oplus})$ and the third-highest RSM overall.  We analyzed data from sectors 31, 42, 43, 70, and 71, using the sliding linear fit detrending method. Our gap/momentum dump trimming algorithm identified 6 regions of heightened noise cut from our analysis: [[2459155.6473, 2459155.8473], [2459447.8051, 2459448.0051], [2459461.3549, 2459461.5549], [2459487.2983, 2459487.3983], [2460253.0648, 2460253.2648], [2460253.7043, 2460253.9043]] BJD.

\begin{figure}
    \centering
    \includegraphics[width=1.0\linewidth]{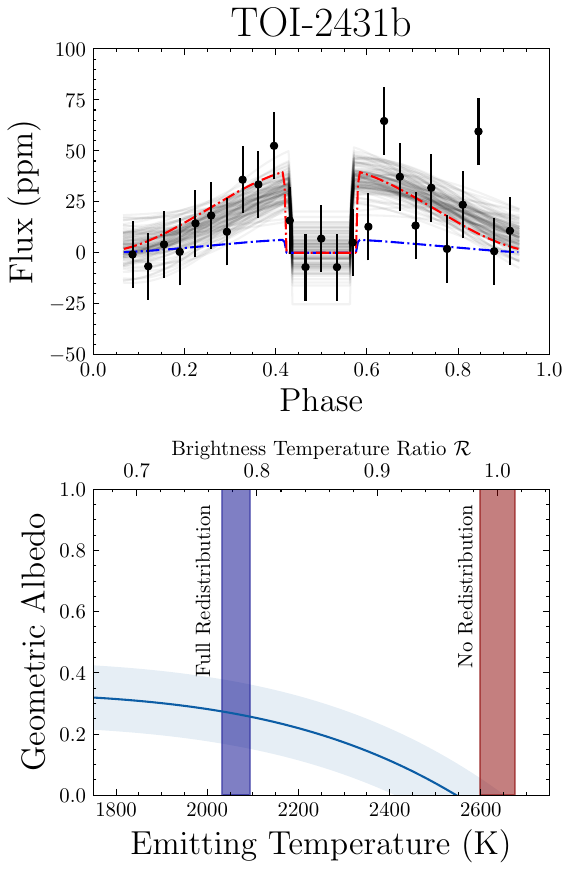}
    \caption{(top) The TESS phase curve of TOI-2431~b, using data from sectors 31, 42, 43, 70, and 71. Grey lines indicate phase curve models created via random draws from the posterior. The red line indicates the thermal emission signal expected in the TESS bandpass for a maximally hot surface ($\mathcal{R}=1$), whereas the blue line shows the expected emission assuming $\mathcal{R}=0.78$, similar to other lava worlds observed in thermal emission.
    (bottom) The TESS-bandpass geometric albedo implied by TOI-2431~b's dayside flux as a function of its emitting temperature.  TOI-2431~b's TESS dayside flux of $35.3^{+10.9}_{-11.2}$\,ppm is well-explained by either reflected light with a moderate geometric albedo of $\sim0.32$ or thermal emission from a maximally hot, dark surface (red).  Scheduled JWST MIRI LRS observations will provide stringent constraints on the emitting temperature and thus geometric albedo.}
    \label{fig:TOI2431b}
\end{figure}

Our first fit enforcing zero nightside flux (the maximum possible thermal nightside flux is 6.1\,ppm)  finds a dayside flux of $33.9^{+10.5}_{-9.9}$\,ppm.  When including free nightside flux, we detect a phase curve signal in the TESS data with a dayside flux of $35.3^{+10.9}_{-11.2}$\,ppm (3.2\,$\sigma$ greater than zero, Figure \ref{fig:TOI2431b}).  The nightside flux of $3\pm11$\,ppm is consistent with zero.  Fits using the eclipse alone (+30 minutes of baseline on each side) retrieve a similar eclipse depth of $32.6^{+11.1}_{-10.7}\,$ppm. The phase curve shows a westward phase offset of $20^{+17}_{-14}$$\degree$ that is $1.4\,\sigma$ consistent with zero.  In Figure \ref{fig:TOI2431b}, we also show the expected contribution from thermal emission as a function of the planet's dayside emitting temperature or `brightness temperature ratio' assuming the planet emits as a blackbody. The brightness temperature ratio $\mathcal{R}$ (e.g., \citealt{xue2024jwst,coy2025population}) is defined as the observed dayside brightness temperature compared to that expected of a zero albedo blackbody, 
\begin{equation}
\mathcal{R}\equiv\frac{T_{day}}{T_{max}}=\left(\frac{2}{3}\right)^{-1/4}(1-A_{eff})^{1/4}\left(\frac{2}{3}-\frac{5}{12}\varepsilon\right)^{1/4} ,
\end{equation}
where,

\begin{equation}
    T_{max}=T_{\star}\sqrt{\frac{R_{\star}}{a}}\left(\frac{2}{3}\right)^{1/4}.
\end{equation}
$\varepsilon$ is the heat redistribution efficiency \citep{cowan2011model}, $A_{eff}$ is the effective albedo, $a$ is the planet's semi-major axis, and $T_{\star}$ and $R_{\star}$ are the effective temperature and radius of the host star. A value of $\mathcal{R}$ near 1 indicates a planet with a low albedo and little heat redistribution, i.e. a `dark bare rock'. Values significantly less than one could indicate efficient heat redistribution or a high Bond albedo, suggesting the presence of an atmosphere.

If the brightness temperature ratio is similar to other lava worlds observed in thermal emission ($\mathcal{R}\sim0.6-0.8$, see \citealt{Coy2026}), then most of the TESS signal is due to reflection. The dayside flux implies a maximum geometric albedo of $0.32\pm0.11$, similar to the Kepler-bandpass geometric albedo inferred for K2-141~b ($0.29^{+0.05}_{-0.06}$, \citealt{zieba2022k2}).  However, due to the planet's extreme instellation, the dayside flux is also well explained by a maximally hot, low albedo surface emitting at a temperature near $T_{max}$, the temperature expected for a zero-heat-redistribution, zero-albedo blackbody.  

While we cannot uniquely constrain the contribution of the planet's thermal emission to the TESS lightcurve, the planet has been observed in a full-orbit phase curve with MIRI LRS as part of JWST GO 8864 (PI: Dang).  This target is a prime candidate for follow-up visible light observations with other space-based telescopes.

\subsubsection{K2-141~b}

K2-141~b is a relatively well-studied ultra-short period lava world.  The Kepler and Spitzer IRAC 4.5$\,\mu$m phase curves were analyzed by \citet{zieba2022k2}, revealing a moderate geometric albedo of $A_{g,\mathrm{K2}}=0.28\pm0.07$ and a low brightness temperature ratio of $\mathcal{R}=0.76\pm0.13$, suggesting a moderately-reflective atmosphere.

Here, we analyze data from TESS sectors 42, 70, and 92. Sector 42 had extremely high amounts of correlated noise compared to typical PDCSAP data.  Inspecting the SAP data for this sector showed that the SPOC pipeline detrending process erroneously introduced high levels of noise.  For our fits, we thus used the SAP data for Sector 42 while continuing to use PDCSAP for sectors 70 and 92.  The SAP flux is adjusted by an estimated dilution factor of 0.9998 and trimmed to use the same timespans as the sector 42 PDCSAP fluxes. We masked transits of the grazing companion K2-141~c assuming a transit duration of 1.2 hours.  Our gap/momentum dump trimming algorithm identified [[2459453.2439, 2459453.4439], [2460232.3740, 2460232.7430], [2460814.0357, 2460814.2517], [2460817.1739, 2460817.3418], [2460827.7800, 2460828.0029]] as regions of heightened noise that we cut from our analysis.  We used a SHO GP kernel for detrending, as the large amounts of stellar variation ($\sim1.5\%$ peak-to-trough) seen in Sector 42 are not well-fit by the pre-fit detrending models.  The SHO kernel also shows better performance than the M\'atern-3/2 and rotation GP kernels, with a $\Delta$BIC (Bayesian Information Criterion) of -8 and -61, respectively.

\begin{figure}
    \centering
    \includegraphics[width=1.0\linewidth]{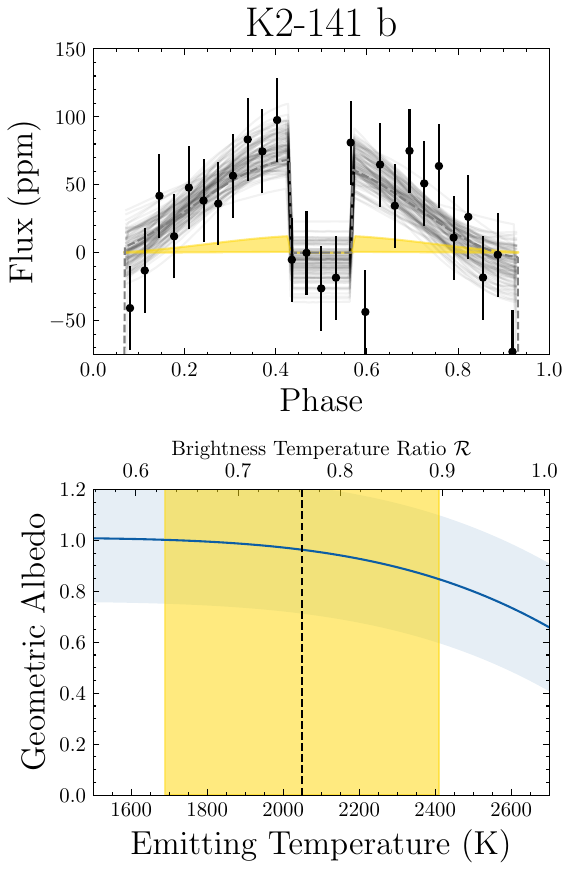}
    \caption{(top) The TESS phase curve of K2-141~b using data from sectors 42, 70, and 92, which shows a very high dayside flux of $79.7^{+19.9}_{-19.8}$\,ppm.  The dashed grey line is the best-fit full phase curve model.  The yellow shaded area represents the expected thermal contribution based on the $\pm1\,\sigma$ constraints on the Spitzer IRAC Channel 2 (4.5$\,\mu$m) brightness temperature reported in \citet{zieba2022k2}. (bottom) The geometric albedo required to reproduce our retrieved dayside flux as a function of emitting temperature.  The intersection of the blue and yellow shaded regions indicate the constraints from the combined TESS and Spitzer observations, which we constrain as $A_{g,TESS}=0.92^{+0.27}_{-0.26}$.}
    \label{fig:K2141}
\end{figure}

Our initial fit enforcing zero nightside flux (\citealt{zieba2022k2} give a 2$\,\sigma$ upper limit of 0.8\,ppm in the TESS bandpass) finds a dayside flux of $80.2^{+18.8}_{-18.3}$\,ppm. When allowing for free nightside flux (shown in Figure \ref{fig:K2141}), we detect the phase curve signal of K2-141~b, retrieving a dayside flux of $79.7^{+19.9}_{-19.8}$\,ppm ($4.0\,\sigma$ from zero) with an eastward phase offset of $14^{+12}_{-11}$$\degree$.  The nightside flux ($-2\pm20\,$ppm) is consistent with zero.  Fits using the eclipse alone (+30 minutes of baseline on each side, excluding sector 42 due to the large rotational signal and using the sliding linear fit for detrending) retrieve a depth of $52\pm24$\,ppm consistent to $1\,\sigma$.

The TESS dayside flux is noticeably larger than the $26.4^{+3.5}_{-2.5}$\,ppm eclipse depth reported for the bluer Kepler bandpass \citep{zieba2022k2}.  Indeed, our dayside flux implies a much higher geometric albedo of $A_{g,\rm TESS}=0.92^{+0.27}_{-0.26}$ compared to the Kepler-bandpass value of $A_{g,\rm Kepler}=0.29^{+0.05}_{-0.06}$.  This may be due to stronger atmospheric emission features in the slightly redder TESS bandpass caused by strong thermal inversions. For example, TiO has strong features over the 0.5-1.0$\,\mu$m range that could cause a deeper eclipse \citep{van2026sensitivity}.  However, due to the lower precision in TESS, the geometric albedo is also $2.3\,\sigma$ consistent with the Kepler value.  Phase curve observations with JWST NIRSpec G395H (JWST GO 2159: PI Espinoza) and MIRI LRS (JWST GO 2347: PI Dang) will better constrain the planet's atmospheric composition and Bond albedo.

\subsubsection{55~Cancri~e}

55~Cancri~e \citep{mcarthur2004detection} is one of the most well-studied ultra-short period planets.  Despite its small planet-to-star radius ratio, its host star's brightness ($m_{Gaia}=5.73$) makes it a prime target for thermal emission and reflection studies, and it has the second-highest RSM of small planets after LTT~9779~b.   Secondary eclipse observations with JWST MIRI and NIRCam have suggested a carbon-rich atmosphere \citep{hu2024secondary}.  However, it is uncertain whether this atmosphere is thick or thin, and four other observations with NIRCam have shown intense, sub-week timescale variability and variable evidence for carbon species \citep{patel2024jwst,snellen2026strong}.

The sector 21, 44, and 46 eclipses were analyzed previously by \citet{valdes2022weak}, finding weak evidence for variability between sectors (see Section \ref{sec:variability} for a discussion on variability) and an average eclipse depth of $8\pm2.5$\,ppm.  Here we add a new sector of TESS data (72) as well as fitting for the full phase curve shape. Our gap trimming algorithm showed two regions of heightened noise cut from our data: [[2458897.0522,2458897.4172], [2459500.7298,2459501.0488]] BJD. We used the sliding linear fit detrending, which showed better performance than the moving median.

\begin{figure}
    \centering
    \includegraphics[width=1.0\linewidth]{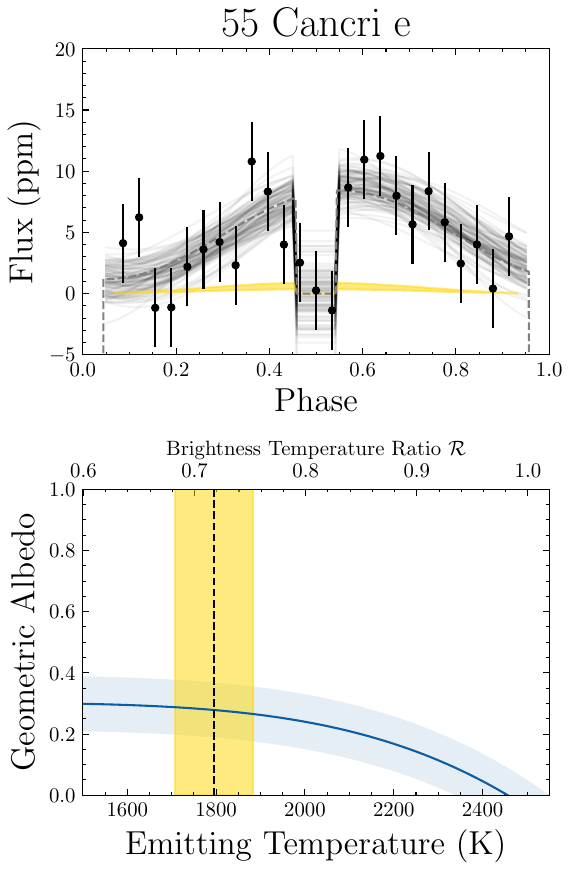}
    \caption{The TESS phase curve of 55~Cancri~e using data from sectors 21, 44, 46, and 72. The dayside flux of $8.1\pm2.4\,$ppm implies a moderate geometric albedo of $0.28\pm0.09$. The thermal component (yellow) is based on the MIRI LRS (5--12$\,\mu$m) brightness temperature reported in \citet{hu2024secondary}.}
    \label{fig:55cnce}
\end{figure}

Our first fit enforcing zero nightside flux (\citealt{mercier2022revisiting} give a 2$\,\sigma$ upper limit of 0.3\,ppm in the TESS bandpass) finds a dayside flux of $7.3\pm1.9$\,ppm. When allowing for free nightside flux (shown in Figure \ref{fig:55cnce}), we detect a dayside flux of $8.1\pm2.4\,$ppm ($3.4\,\sigma$ from zero).  The retrieved nightside flux of $1.2^{+2.4}_{-2.3}$\,ppm is consistent with zero.  Fitting the eclipse alone with the same baseline prescription as \citet{valdes2022weak} (the eclipse plus 3.65 hours on each side) gives an eclipse depth of $6.2\pm2.3\,$ppm, $1\,\sigma$ consistent with the full phase curve.

The data also show tentative evidence for a westward phase offset of $25^{+14}_{-13}$$\degree$. CHEOPS also observed 29~individual phase curves of 55~Cancri~e \citep{valdes2023investigating}. These observations also show some evidence for an average westward visible phase offset (weighted average of $14\pm3\degree$ using the values reported in \citealt{valdes2023investigating}), but the scatter between individual observations is very high.  Combined with the strong eastward offset of $41\pm12\degree$ reported for Spitzer thermal infrared phase curve observations \citep{demory2016map} (although this has been contested, see \citealt{mercier2022revisiting}), this may suggest that reflective clouds preferentially form on the cooler westward side of the planet, or that they are being advected from the nightside through eastward jets, as suggested for LTT~9779~b \citep{coulombe2025highly}.

Our retrieved dayside flux is comparable to the average eclipse depth reported for CHEOPS observations of 55~Cancri~e ($12\pm3$\,ppm, \citealt{demory202355}), which covers a slightly bluer wavelength range where geometric albedo could be larger.  Future NIRISS eclipse observations with JWST (GO~9825, PI Hoeijmakers) may better constrain 55~Cancri~e's visible/NIR geometric albedo and potential variability.

\subsubsection{TOI-561~b}

TOI-561~b \citep{weiss2021tess} is a low-density ultra-hot lava world ($T_{irr}=3300\,$K, $R=1.4\,R_{\oplus}$).  The planet's low density, combined with a low dayside temperature measured by JWST NIRSpec suggests a thick atmosphere \citep{teske2025thick,boucher2026a}.  However, as no molecular features have been detected, the atmosphere's composition is unknown.

The TESS sectors 8, 34, 45, and 46 secondary eclipse of TOI-561~b was previously analyzed by \citet{patel2023cheops}, who retrieved an eclipse depth of $27\pm11$\,ppm.  This is much higher than the thermal component expected from the JWST-retrieved brightness temperature, which should be $<$1~ppm in the TESS bandpass.  We add data from one more TESS sector (Sector 72) and fit for the dayside flux using the full phase curve.  We mask the transits of the other planets in the system (c, d, and e) using the median-fit parameters reported in \citet{piotto2024architecture}, which were fit using all currently-available TESS data.  The other planets' orbital periods are sufficiently long ($>10.8\,$d) compared to TOI-561~b (0.45\,d) that any potential phase curve signal should be removed with our detrending algorithms.  The maximum reflected signal of the largest potential contaminant, TOI-561~c, is 1.9\,ppm, whereas the expected thermal signal is $<0.001\,$ppm.  Our gap/momentum dump trimming algorithm identified 3 regions of heightened systematics that we cut from our analysis: [[2458535.2274, 2458535.3274], [2459266.6021, 2459266.9305], [2460275.9327, 2460275.9577]] BJD. We use the sliding linear fit detrending algorithm, which showed slightly better performance than the moving median detrending and showed no signs of residual correlated noise.

\begin{figure}
    \centering
    \includegraphics[width=1.0\linewidth]{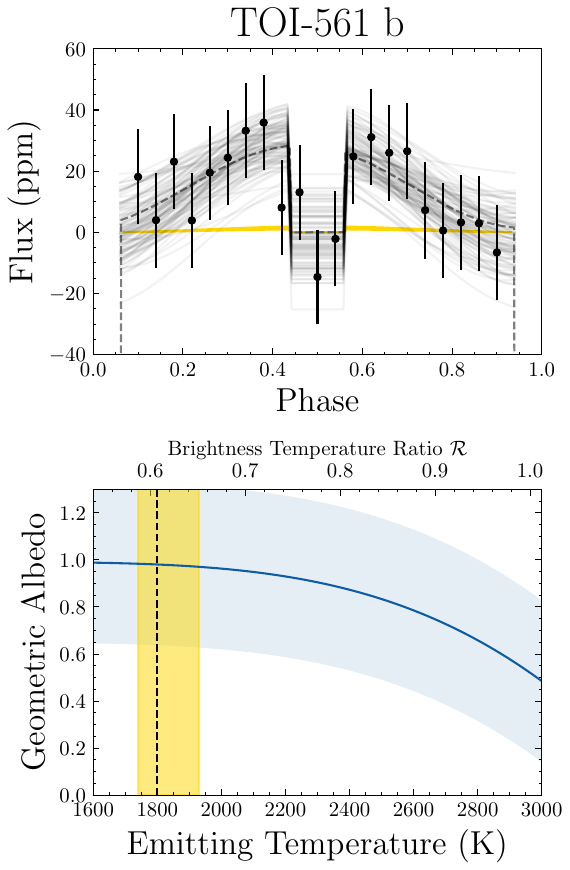}
    \caption{The TESS phase curve of TOI-561~b, using data from sectors 8, 35, 45, 46, and 72.  The expected thermal contribution (yellow) is based on the JWST NIRSpec/G395H dayside brightness temperature reported in \citet{teske2025thick}. The TESS dayside flux of $31.8^{+11.2}_{-10.9}\,$ppm  implies an extremely high geometric albedo of  $0.98^{+0.36}_{-0.35}$, similar to the high Bond albedo of $0.74^{+0.08}_{-0.10}$ derived from JWST observations \citep{boucher2026a}.}
    \label{fig:toi561}
\end{figure}

Our first fit enforcing zero nightside flux (\citealt{boucher2026a} give a 2$\,\sigma$ upper limit of 0.04\,ppm in the TESS bandpass) finds a dayside flux of $30.6\pm10.1$\,ppm.  When allowing for a free nightside flux (shown in Figure \ref{fig:toi561}), we detect a dayside flux of $31.8^{+11.2}_{-10.9}\,$ppm (2.9$\,\sigma$ from zero).  The nightside flux is consistent with zero at $4\pm11\,$ppm. Our eclipse-only analysis using the same baseline as \citet{patel2023cheops} corroborates their results, retrieving an eclipse depth of $28\pm10$\,ppm. 

This dayside flux implies an extremely high geometric albedo of $0.98^{+0.36}_{-0.35}$, similar to that inferred for K2-106~b ($A_{g}=0.90\pm0.26$, \citealt{singh2022probing}), a lava world at a similar temperature ($T_{irr}=3299\,$K).  This high geometric albedo, if confirmed, would be the highest ever measured for an exoplanet, and would imply that its atmosphere is strongly backscattering. Such an ultra-reflective atmosphere would be consistent with recent analysis of the full JWST/NIRSpec phase curve, which suggests a very high Bond albedo of $A_{B}=0.74^{+0.08}_{-0.10}$ \citep{boucher2026a}. Accounting for these high albedos may require aerosol compositions/morphologies or atmospheric processes that are not included in current models: \citet{boucher2026a} showed that current models predict that the contribution of possible condensates (including MgSiO$_3$ and SiO$_2$) to the Bond albedo is only $\sim0.22$.  This is because TOI-561~b is so hot that clouds only form in cooler regions away from the substellar point.

\subsubsection{TOI-1444~b}

TOI-1444~b \citep{dai2021tks} is a $R=1.4\,R_{\oplus}, M=3.9\,M_{\oplus}, T_{irr}=3010\,\mathrm{K}, P=0.47$\,d lava world at an irradiation temperature similar to TOI-561~b and is near the TESS continuous viewing zone.  The TESS sectors 15, 16, 17, 18, 19, 22, 24, 25, and 26 phase curve was analyzed previously in \citet{dai2021tks}, where they found tentative evidence for a dayside flux of $27\pm12$\,ppm.  This would imply either an extremely hot surface or a high geometric albedo near one ($F_{ref,max}/F_{\star}\approx27\,$ppm).  Here, we add new data from sectors 49, 52, 56, 57, 58, 59, 60, 73, 76, 77, 78, 79, 84, 85, and 86. Our gap trimming algorithm identified several regions of heightened noise cut from our analysis, which constituted $\sim2\%$ of the data. 

\begin{figure}
    \centering
    \includegraphics[width=1.0\linewidth]{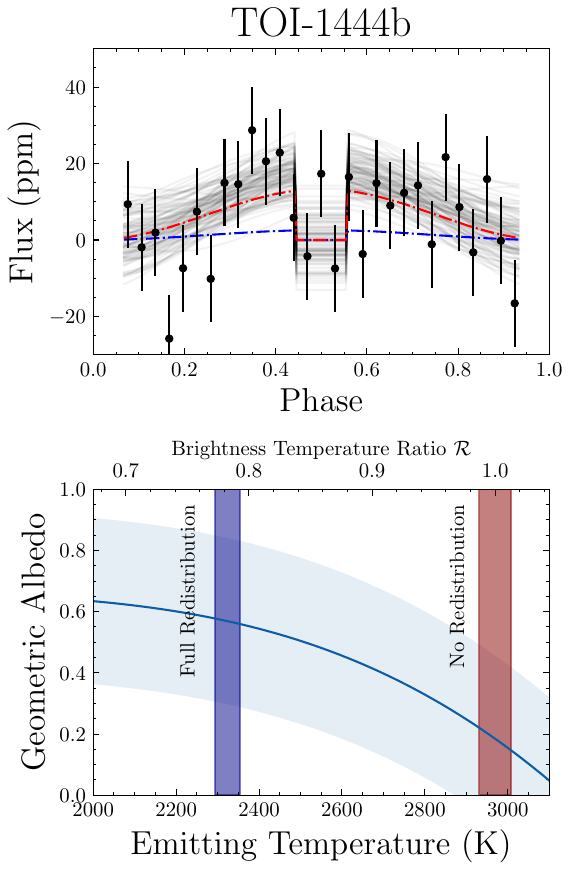}
    \caption{The TESS phase curve of TOI-1444~b using data from sectors 15--19, 22, 24--26, 49, 52, 56--60, 73, 76--79, and 84--86. The red line indicates the thermal emission signal expected in the TESS bandpass for a maximally hot surface ($\mathcal{R}=1$), whereas the blue line shows the expected emission assuming $\mathcal{R}=0.78$, similar to other lava worlds observed in thermal emission. The dayside flux of $17.7^{+7.3}_{-7.2}\,$ppm implies either a high geometric albedo of $0.57^{+0.29}_{-0.28}$ or a maximally hot surface, which predicts a dayside flux ($13.8^{+2.7}_{-1.9}$\,ppm) 1$\,\sigma$ consistent with our constraint.}
    \label{fig:toi1444b}
\end{figure}

Our initial fit enforcing zero nightside flux (the maximum possible thermal nightside flux is 2.4\,ppm) finds a dayside flux of $20.4^{+6.4}_{-6.3}\,$ppm.  When allowing for free nightside flux (shown in Figure \ref{fig:toi1444b}), we detect a dayside flux of $17.7^{+7.3}_{-7.2}\,$ppm (2.5$\,\sigma$ from zero), consistent with but improving upon the precision of \citet{dai2021tks}.  The retrieved nightside flux is consistent with zero at $-5^{+8}_{-7}\,$ppm.

Fitting the eclipse alone (plus 1 hr of baseline on each side) retrieves an eclipse depth of $16.4^{+7.5}_{-7.7}\,$ppm, 2.2$\,\sigma$ from zero and 1$\,\sigma$ consistent with the full phase curve fit.  Shown in Figure \ref{fig:toi1444b}, this high dayside flux implies a high geometric albedo of $0.57^{+0.29}_{-0.28}$ but is also consistent with a maximally hot surface, which predicts a dayside flux of $13.8^{+2.7}_{-1.9}$\,ppm.  Future mid-infrared secondary eclipse observations with JWST may help break this degeneracy.

\subsection{Tentative Detection}

We also found one high-RSM target with tentative evidence (2--3$\,\sigma$) for dayside flux with TESS.  Data from future sectors may be able to improve the confidence of this detection to $\geq3\,\sigma$. As this detection is tentative, we do not report the implied geometric albedo.

\subsubsection{TOI-332~b}
TOI-332~b \citep{osborn2023toi} is a very dense (9.6\,g/cm$^3$) ultra-hot Neptune. We fit for the TESS phase curve using data from sectors 2, 3, 28, 68, 95, 96, 105, and 106. The sector 2-3 data (only available at 30 minute cadences) were supersampled to a resolution of 2 minutes.  We used the moving median method for detrending and saw no evidence for residual correlated noise.

We retrieve a dayside flux of $80.1\pm29.7\,$ppm (2.7$\,\sigma$ from zero), high compared to the maximum reflected light signal of 73\,ppm and maximum thermal signal (assuming $\mathcal{R}=1$) of 22\,ppm. The nightside flux of $40\pm30\,$ppm is slightly high, but 1.3$\,\sigma$ consistent with zero.  Fitting for the eclipse depth using a baseline of two hours on each side yields $84\pm31\,$ppm.  Future TESS data could help confirm or reject this tentative signal, requiring $\sim$2 more sectors if the true dayside flux is 80.1\,~ppm. 

\subsection{Informative Non-Detections}
For some targets where we did not confidently detect a dayside flux, the retrieved dayside flux is precise enough to place informative upper limits on their geometric albedos. These upper limits may help constrain model predictions of the reflected light spectra of lava worlds.

\subsubsection{TOI-431~b}
TOI-431~b  is an ultra-short period lava world ($T_{irr}=2600$\,K, $R=1.28\,R_{\oplus}$, $P=0.49\,$d), that has been observed in secondary eclipse by both Spitzer \citep{monaghan2025low} and JWST \citep{Smith2026}, revealing a dayside brightness temperature ratio of $\mathcal{R}=0.81\pm0.10$, suggesting an atmosphere.  The planet's full phase curve has also been observed with JWST (GO 8864, PI: Dang).
    
Its short period and bright host star ($m_{gaia}=8.79$) makes it a prime candidate for reflected light studies, with an RSM of 0.85.  Here we analyze 4 sectors of TESS data, masking out transits of the companion TOI-431~d using the median-fit parameters derived in \citet{Smith2026}.  Our gap trimming algorithm identified several regions of heightened noise that we cut from our analysis.  The majority of these are in sector 98, which shows strong evidence for earthshine.  We used sliding linear fit detrending, retrieving a dayside flux of $-2\pm7\,$ppm and a nightside flux of $-1\pm7\,$ppm.  The corresponding 2\,$\sigma$ (97.7\%) upper limit on the dayside flux is 10.9\,ppm. TOI-431~b has a maximum reflected light signal of 23.3~ppm assuming $A_{g}=1$ and the thermal component in TESS derived from JWST MIRI should be $\sim1.3$~ppm.  We thus infer that the geometric albedo of TOI-431~b is $A_{g,TESS}<0.41$ to $2\,\sigma$ confidence.

\subsubsection{GJ~367~b}

GJ~367~b \citep{lam2021gj} is a dense (6.9\,g/cm$^3$) sub-Earth orbiting a bright ($m_{Gaia}=9.15$) M dwarf.  The planet was observed over a full phase curve JWST/MIRI LRS \citep{zhang2024gj}, revealing a dayside temperature near the `dark bare rock' expectation ($\mathcal{R}\approx1$).  Here, we analyzed data from sectors 9, 35, 36, 63, 89, 90, and 99 using the moving median detrending algorithm. We used SAP data for sectors 63 and 99, as the PDCSAP data showed higher amounts of anomalous correlated noise.  The data are trimmed to the same timespans as PDCSAP and use dilution factors of 0.8812 and 0.9019, respectively.   Our gap trimming algorithm identified several regions of heightened noise cut from our analysis, which constituted $\sim5\%$ of the data.

We retrieve a low dayside flux of $0\pm6$\,ppm, with a 2$\,\sigma$ upper limit of 11.1\,ppm. The nightside flux is consistent with zero ($-5\pm6\,$ppm).  The thermal flux propagated from the MIRI-derived dayside temperature should be $\approx1.6$\,ppm in the TESS bandpass, allowing us to place an informative $2\,\sigma$ upper limit of $A_{g}<0.46$, consistent with the `dark bare rock' inference of JWST MIRI observations.

\subsection{Comparison to Photon-Limited Noise}

Is there more room for improvement on the precision of our phase curve parameters given the current amount of TESS data?  To answer this question, we compare the posterior constraints from \texttt{SPARTESS} with the precision that (for the same cadences) photon statistics alone would predict. Rather than using a population noise model as a function of the TESS magnitude, we compute the photon-limited precision of each star directly from its own data. Pre-launch noise model predictions  (e.g., \citealt{Sullivan2015,Stassun2018}) are known to be pessimistic for brighter stars: they overstate the noise floor of bright stars because of the worst-case 60\,ppm\,hr$^{-\frac{1}{2}}$ systematic term they assume \citep{kunimoto2022predicting}. For several of our targets, the prediction of \citet{Stassun2018} over-predicts the observed white noise scatter by 10--20\%, so a comparison against it would misleadingly suggest that our fit precisions are better than photon-limited.

We compute the photon noise of each star from the electron count reported in the SPOC aperture, where $\sigma_{\rm phot} = 1/\sqrt{N_\star}$ per cadence, where $N_\star$ is the \texttt{SAP\_FLUX} electrons per second flux multiplied by the 95.04\,s effective integration time per two-minute cadence.   Adding the sky background collected in the aperture (\texttt{SAP\_BKG}) and the read noise of its pixels over the 48 reads gives the full instrumental floor, $\sigma_{\rm phot+sky+read} = \sqrt{N_\star + N_{\rm sky} + n_{\rm pix}\,n_{\rm read}\,\sigma_{\rm read}^2}\,/\,N_\star$, which exceeds the pure photon noise by 5--10\% for the unsaturated stars. TESS 2 minute cadences, unlike the up-the-ramp sampling of JWST, are a destructive co-adding of $n_{\rm read}=48$ separate frames of 1.98\,s exposures, each introducing their own independent read noise.

Separately, we measure the white noise actually present in the light curve, $\sigma_{\rm meas}$, as the scatter of consecutive-cadence differences divided by $\sqrt{2}$. This measurement is independent of the choice of detrending model, which could bias the scatter of the best-fit residuals, and it sets the best precision that any analysis can extract from the given data. Because $F_\star$, $F_{n}$, and $D_{1}$ are free parameters, the eclipse alone separates $F_d$ from $F_n$, so the precision depends on the in-eclipse time and an extrapolation of the out-of-eclipse baseline to mid-eclipse phase, where approximately
\begin{equation}
    \sigma(F_d) \simeq \sigma\sqrt{1/T_{\rm ecl} + 3/T_{\rm out}},
    \label{eq:sigfd}
\end{equation}
with $T_{\rm ecl}$ and $T_{\rm out}$ being the total time spent in and out of eclipse and $\sigma$ being any of the noise estimates above.  The factor of 3 arises because the baseline is a sum of a constant (which contributes $\sigma^{2}/T_{\rm out}$ variance) and cosine amplitude (which contributes $2\sigma^{2}/T_{\rm out}$) fitted to the out-of-eclipse data. The values in Table~\ref{tab:photon} propagate each noise level through the full covariance of the linear model with $F_\star$, $F_d$, $F_n$ and $D_1$ free. The approximation of Equation \ref{eq:sigfd} reproduces these values to within 8\%.

\begingroup
\setlength{\tabcolsep}{3pt}
\begin{deluxetable*}{llcccccccccc}
\tabletypesize{\footnotesize}
\tablecaption{Photon-limited, White Noise-limited, and Achieved Precision of TESS Dayside Fluxes\label{tab:photon}}
\tablehead{
\colhead{Planet} & \colhead{Detrending} & \colhead{$m_{\rm TESS}$} & \colhead{$T_{\rm tot}$} & \colhead{$T_{\rm ecl}$} & \colhead{$\sigma_{\rm phot+sky+read}$} & \colhead{$\sigma_{\rm meas}$} & \colhead{$\sigma_{\rm meas}/\sigma_{\rm p+s+r}$} &
\multicolumn{3}{c}{$\sigma(F_d)$ (ppm)} \\
\colhead{} & \colhead{} & \colhead{} & \colhead{(hr)} & \colhead{(hr)} & \colhead{(ppm hr$^{-\frac{1}{2}}$)} & \colhead{(ppm hr$^{-\frac{1}{2}}$)} & \colhead{} &
\colhead{photon limit}& \colhead{white noise limit} & \colhead{fit}
}
\startdata
55~Cnc~e     & Sanchis    &  5.21 & 1976 & 170 &   16.5 &  26.1 & 1.58 &   1.5 &  2.3 &  2.4 \\
LTT~9779~b   & Sanchis    &  9.10 & 1887 &  65 & 106.9 & 116.6 & 1.09 &  14.7 & 16.0 & 16.0 \\
TOI-2431~b   & Sanchis    &  9.52 & 2349 & 322 &  132.9 & 148.5 & 1.12 &   9.7 & 10.8 & 11.0 \\
TOI-561~b    & Sanchis    &  9.53 & 2256 & 278 & 133.7 & 148.3 & 1.11 &   10.1 & 11.2 & 11.2 \\
K2-141~b     & GP, SHO    & 10.03 & 1102 & 151 &  177.2 & 193.3 & 1.09 &  18.8 & 20.5 & 20.6 \\
WASP-121~b   & GP, M\'atern & 10.06 & 3088 & 259 &  180.7 & 206.6 & 1.14 & 15.1 & 17.2 & 18.5 \\
TOI-1444~b  & Sanchis & 10.15 & 10650 & 1180 & 187.1 & 205.7 & 1.10 &   6.7 & 7.3 &  7.4 \\
\enddata
\tablecomments{ $T_{\rm tot}$ is the total duration of the fitted cadences, and $T_{\rm ecl}$ the total in-eclipse-time. $\sigma_{\rm phot+sky+read}$ is the photon noise of the star from its electron counts adding the sky background and read noise of the aperture, and $\sigma_{\rm meas}$ is the measured point-to-point scatter of the raw data used in the fits--both are reported per one-hour bin. The `photon limit' and `white noise limit' columns are the expected precision of the dayside flux assuming white noise based on the expected photon limit and observed scatter, respectively, whereas `fit' is the precision achieved in our \texttt{SPARTESS} fits. All fit precisions are very close to the white noise limit, indicating that there is no additional precision to be gained from the current data. 55~Cnc, which is significantly noisier than the photon noise prediction, is saturated in the TESS cameras and its noise is dominated by pointing jitter and bleed-column effects.
}
\end{deluxetable*}
\endgroup

Table \ref{tab:photon} compares the photon-limited, instrumental floor, and white noise-limited precision for each of our fits where we detected the dayside flux at $>\,3\sigma$ (we use results from the free nightside flux fits).  We also report the predicted white-noise limited precision on the dayside flux for each of these scenarios. The measured scatter is typically within $\sim20\%$ of the pure photon limit and $\sim15\%$ of the instrumental limit including read noise and sky background, except for 55 Cancri where the TESS cameras are saturated and the noise is dominated by pointing jitter and bleed effects.   Our fits all reach precision within $\sim 4\%$ of the empirically measured white noise-limited precision, indicating that our precision is limited by the data quality, not our detrending and fitting algorithms.  The one exception is WASP-121~b at 8\% higher than the white noise limit, which shows high amounts of correlated noise with timescales of $0.15\sim0.37$\,d that share power with the 1.27\,d phase curve.  We conclude that there is little-to-no further precision to be gained from the existing TESS data, and tighter constraints would require additional sectors, or a reduction of model freedom such as fixing $F_n$ to zero.

\section{Discussion and Interpretation} \label{sec:discussion}

\subsection{Determining Geometric Albedos}

\begingroup
\setlength{\tabcolsep}{4pt}
\begin{deluxetable*}{lcccccccc}
\tabletypesize{\footnotesize}
\tablecaption{Host Star and System Parameters Used in Geometric Albedo Nested Sampling}
\tablewidth{0pt}
\tablehead{
\colhead{Planet/Star} & \colhead{$T_{\rm eff,\star}$ (K)} & \colhead{log($g[cm/s^{2}]$)} & \colhead{[M/H]} & \colhead{$a/R_{\star}$} & 
\colhead{$R_{p}/R_{\star}$} & Reference}
\startdata
    55 Cnc e & $5172\pm18$ & $4.43\pm0.02$  & $0.35\pm0.10$  & $3.52\pm0.01$  & $0.0182\pm0.0002$ & \citet{bourrier201855} \\
    TOI-2431 b & $4109^{+28}_{-27}$ & $4.68\pm0.05$  & $-0.02\pm0.13$  & $2.08\pm0.05$$^{\dagger}$  & $0.02131^{+0.00033}_{-0.00032}$ & \citet{tacs2026earth} \\
    K2-141 b & $4599\pm79$ & $4.62^{+0.02}_{-0.03}$  & $-0.06^{+0.08}_{-0.10}$  & $2.292^{+0.053}_{-0.060}$  & $	0.02037\pm0.00046$ & \citet{malavolta2018ultra} \\
    Kepler-10 b & $5708\pm28$ & $4.344\pm0.004$ & $-0.15\pm0.04$ & $3.408\pm0.004$ & $0.01254\pm0.00013$  & \citet{dumusque2014kepler} \\
    Kepler-407 b & $5556\pm60$ & $4.45\pm0.03$ & $0.41\pm0.04$ & $3.221\pm0.085$ &	$0.01040\pm0.00028$ &  \citet{fulton2018california} \\
    Kepler-78 b & $5058\pm50$ & $4.577^{+0.022}_{-0.026}$$^{\dagger}$ & $-0.18\pm0.08$ & $2.58\pm0.05$$^{\dagger}$ & $0.01472\pm0.00037$$^{\dagger}$ &\citet{bonomo2023cold}\\
    K2-131 b & $5120\pm71$ & $4.582\pm0.021$$^{\dagger}$ & $-0.04\pm0.07$ & $2.651\pm0.050$$^{\dagger}$ & $0.02060^{+0.00095}_{-0.00084}$$^{\dagger}$ & \citet{bonomo2023cold} \\
    TOI-561 b & $5372\pm70$ & $4.50\pm0.12$  & $-0.40\pm0.05$  & $2.683\pm0.029$  & $0.01519\pm0.00028$ & \citet{lacedelli2022investigating}\\
    TOI-1444~b & $5430\pm90$ & $4.49\pm0.03$ & $0.13\pm0.06$ & $2.730^{+0.078}_{-0.074}$ & $0.01410^{+0.00060}_{-0.00058}$ & \citet{dai2021tks} \\
    K2-106~b & $5578\pm46$ & $4.408\pm0.017$ & $-0.005\pm0.059$ & $2.858\pm0.047$ & $0.01480\pm0.00030$ & \citet{palethorpe2026constraining} \\
\enddata
\tablecomments{Stellar and system parameters used for nested sampling, in order of planet irradiation temperature. $^{\dagger}$These values are derived from reported stellar mass, radius, and planet radius constraints.}\label{tab:hoststars}
\end{deluxetable*}
\endgroup

We use our dayside fluxes reported above to derive estimates of the planets' geometric albedos.  Unlike the simple contour plots above (e.g., Figure \ref{fig:TOI2431b}), here we account for uncertainties in stellar and orbital parameters, which can introduce uncertainty on the geometric albedo \citep{coy2025population,monaghan2026uniform}. The stellar and orbital parameters used in nested sampling are shown in Table \ref{tab:hoststars}.

We follow the methodology of \citet{coy2025population} and simultaneously fit for the visible/NIR and mid-IR eclipse depths/dayside fluxes.  We do this by first calculating the bandpass-integrated eclipse depth from thermal emission for a given instrument via,

\begin{equation} \label{eq:FpFs}
\begin{split}
    \left(\frac{F_{p}}{F_{\star}}\right)_{therm}=\left(\frac{R_{p}}{R_{\star}}\right)^{2}\times \\
    \frac{\int_{\lambda_{min}}^{\lambda_{max}}\frac{\pi B_{\lambda}(\mathcal{R}\times T_{max}(T_{\star},a/R_{\star}))}{hc/\lambda}W_{inst,\lambda}d\lambda}{\int_{\lambda_{min}}^{\lambda_{max}}\frac{M_{\lambda}(T_{\star},\log{(g)},[\mathrm{M}])}{hc/\lambda}W_{inst,\lambda}d\lambda},
\end{split}
\end{equation}
where $B_{\lambda}$ is the blackbody Planck function, [M] is the log$_{10}$ stellar metallicity relative to solar, $\log(g)$ is the stellar surface gravity, $M_{\lambda}$ is the model stellar flux, determined from PHOENIX stellar models \citep{husser13} interpolated using \texttt{pysynphot} \citep{pysynphot}.

We go beyond \citet{coy2025population} by adding a geometric albedo component that only applies to the visible/NIR eclipse depth, i.e.,

\begin{equation}
    \left(\frac{F_{p}}{F_{\star}}\right)_{vis}(\lambda)=A_{g}\left(\frac{R_{p}}{R_{\star}}\times\frac{R_\star}{a}\right)^{2}+ \left(\frac{F_{p}}{F_{\star}}\right)_{therm}(\lambda).
\end{equation}
The geometric albedos of silicate clouds are expected to depend strongly on wavelength, with generally low-to-near-zero geometric albedos at wavelengths longer than $2.5\,\mu$m (e.g., \citealt{marley1999reflected,coulombe2025highly}) probed by thermal emission observations.   Here, we focus only on targets with a dayside flux detection significance of $>3\,\sigma$, which includes K2-141~b, TOI-2431~b, 55~Cancri~e, and TOI-561~b from our TESS analyses, Kepler-10~b, K2-106~b, K2-131~b, Kepler-407~b, and Kepler-78~b from \citet{singh2022probing}, K2-141~b from \citet{zieba2022k2}, and 55~Cancri~e from CHEOPS \citep{demory202355}.  We again use the (less precise) dayside fluxes derived from our free nightside flux fits as a more conservative estimate of the true dayside flux and geometric albedo.

\subsubsection{Kepler Targets and TOI-2431~b/TOI-1444~b} \label{sec:r78}

For most of the planets with Kepler dayside fluxes reported in \citet{singh2022probing} and our TESS dayside fluxes derived for TOI-2431~b and TOI-1444~b, there are not yet mid-infrared measurements of the planets' emitting temperatures.   This makes it harder to determine the geometric albedo, since (as shown in Figure \ref{fig:TOI2431b}) these planets are hot enough to have a significant amount of thermal emission in the visible. As a workaround, we consider two end-member cases when deriving these albedos. 
The first endmember assumes the planet is a maximally hot blackbody with no heat redistribution (i.e., $\mathcal{R}=1$).  This assumption is inherently pessimistic and not energetically consistent--it assumes that the planet reflects no light but a geometric albedo above zero implies some light is reflected. The second endmember assumes $\mathcal{R}(\varepsilon=1)=0.78$. This value is more representative of the retrieved $\mathcal{R}$ values for hot ($T_{irr}>2400\,$K) lava worlds from JWST observations, which currently span $\mathcal{R}=0.57^{+0.08}_{-0.09}$ for HD~3167~b \citep{Coy2026} to $\mathcal{R}=0.81\pm0.10$ for TOI-431~b \citep{Smith2026}.   While this value is calculated assuming full heat redistribution to the nightside, it does not assume a thick atmosphere;  $\mathcal{R}$ can also be lowered by a high Bond albedo without the need for efficient heat redistribution.  We thus plot values from this endmember case in Figure \ref{fig:population}, and the $\mathcal{R}=1$ case in Appendix Figure \ref{fig:populationR1}.   However, note that the $\mathcal{R}=0.78$ assumption does not represent the `maximum geometric albedo,' as there is still some thermal contribution in the visible.  Mid-infrared thermal emission observations with JWST will ultimately be needed to break the thermal emission--reflection degeneracy for these targets.

\begin{deluxetable*}{lcccccccc}[t]
\tablecaption{Planet Parameters and Homogeneously-Derived Geometric Albedos Used in This Study}
\tablewidth{0pt}
\tablehead{
\colhead{Planet} & \colhead{$T_{irr}$} & \colhead{Vis. Dayside Flux} & \colhead{Vis. Instrument} & \colhead{MIR Dayside Flux} & \colhead{MIR Instrument} & \colhead{$A_{g}$} \\[-2mm]
\colhead{} & \colhead{(K)} & \colhead{(ppm)} & \colhead{} & \colhead{} & \colhead{}& \colhead{}}
\startdata
    55~Cancri~e & $2756\pm10$ & $12\pm3$ & CHEOPS & \citet{hu2024secondary} & MIRI/LRS & $0.44\pm0.11$ \\
     & & $8.1\pm2.4$ & TESS &   & & $0.29\pm0.09$ \\ \hline
    TOI-2431~b & $2918\pm42$  & $35.3^{+10.9}_{-11.2}$  & TESS & $\mathcal{R}=1$ & --- & $-0.04\pm0.11$  \\
   & &  & & $\mathcal{R}=0.78$  & --- &  $0.27\pm0.11$\\ \hline
    K2-141~b & $3038\pm65$ & $26.4^{+3.5}_{-2.5}$ & Kepler & \citet{zieba2022k2} & IRAC Ch 2 & $0.29^{+0.05}_{-0.06}$ \\
    & & $79.7^{+19.9}_{-19.8}$& TESS &  &   & $0.92^{+0.27}_{-0.26}$ \\\hline
    Kepler-10~b & $3094\pm23$ & $10.4\pm0.9$ & Kepler & $\mathcal{R}=1$ & --- & $0.55\pm0.07$   \\
    & &  & & $\mathcal{R}=0.78$ & --- &  $0.74\pm0.07$  \\ \hline
     Kepler-407~b & $3096\pm53$  & $6.0^{+1.9}_{-1.6}$ & Kepler & $\mathcal{R}=1$ & --- & $0.33^{+0.19}_{-0.17}$    \\
     &  & &  & $\mathcal{R}=0.78$ & --- & $0.56^{+0.18}_{-0.17}$    \\ \hline
    Kepler-78~b & $3144\pm45$  & $12.4\pm1.3$ & Kepler & $\mathcal{R}=1$ & --- &  $0.14\pm0.05$\\
     &   & &  & $\mathcal{R}=0.78$ & --- &  $0.35\pm0.05$ \\ \hline
     K2-131~b &  $3145\pm53$  & $28\pm9$ & Kepler & $\mathcal{R}=1$ & --- & $0.22^{+0.16}_{-0.15}$    \\
    & & & & $\mathcal{R}=0.78$ & --- &  $0.43^{+0.16}_{-0.15}$   \\ \hline
    TOI-561~b & $3280\pm48$ & $31.8^{+11.2}_{-10.9}$ & TESS  & \citet{boucher2026a} & NIRSpec/G395H & $0.93\pm0.32$ \\ \hline
    TOI-1444~b & $3285\pm70$  & $17.7^{+7.3}_{-7.2}$ & TESS & $\mathcal{R}=1$ & --- & $0.17^{+0.30}_{-0.28}$  \\
   & &  & & $\mathcal{R}=0.78$  & --- & $0.57^{+0.29}_{-0.28}$ \\ \hline
    K2-106~b & $3299\pm 39$ & $25.3^{+7.7}_{-7.6}$ & Kepler & $\mathcal{R}=1$ & --- & $0.67\pm0.29$    \\
     &  &  &  & $\mathcal{R}=0.78$ & --- &  $0.90\pm0.29$   \\ \hline
\enddata
\tablecomments{The CHEOPS dayside flux is taken from the averaged eclipse depth reported in \citet{demory202355}. Dayside fluxes for the Kepler targets are taken from \citet{singh2022probing}, except for K2-141~b where we use \citet{zieba2022k2}.  Planets without mid-IR constraints on their dayside emitting temperatures use two end-member assumptions for their brightness temperature ratios (see Section \ref{sec:r78}) to calculate the thermal contribution in the given visible band.
}\label{tab:Rvalues}
\end{deluxetable*}

\subsection{Comparison to Hot Jupiters}

We compare our geometric albedos to values for hot Jupiters compiled in \citet{heng2026albedo} (and references therein), using data from Kepler, TESS, CHEOPS, COROT, and JWST/NIRISS. For simplicity we ignore HD~189733~b, which shows a large discrepancy between its HST (0.29-0.45$\,\mu$m) and CHEOPS-derived geometric albedos ($0.40\pm0.12$ and $0.076\pm0.016$, respectively). We also include geometric albedos derived for LTT~9779~b from JWST/NIRISS \citep{coulombe2025highly}.  For LTT~9779~b CHEOPS data, we use the eclipse depth derived in \citet{saha2025high} ($89.9\pm13.7\,$ppm) to re-derive a geometric albedo correctly accounting for the thermal contribution in the CHEOPS band constrained by the JWST NIRSpec+NIRISS-derived dayside temperature reported in \citet{ashtari2026heat} ($2252^{+19}_{-17}$\,K), retrieving a value of $0.53\pm0.11$.  Note that this simple model does not include atmospheric spectral features, but is now more consistent with the NIRISS-derived value of $0.50\pm0.07$.

We compare geometric albedo measurements for two populations: small or sub-Saturn planets ($R<5\,R_{\oplus}$) and hot Jupiters with $R>7\,R_{\oplus}$. The differences are large (Figure \ref{fig:population}).  Hot Jupiters tend to have low geometric albedos near that of the airless bodies of the Moon and Mercury ($\sim0.14$, \citealt{mallama2017albedos}), whereas lava worlds show both a broader distribution and generally higher geometric albedos closer to that of Venus (0.69).  The median and 1$\,\sigma$ standard deviations of each distribution are $A_{g}=0.49^{+0.34}_{-0.20}$ for the small planets and $A_{g}=0.10^{+0.12}_{-0.06}$ for the hot Jupiters. While we assume $\mathcal{R}=0.78$ for some of the small planets without thermal emission measurements, assuming a maximally hot surface ($\mathcal{R}=1$, which is overly pessimistic due to their non-zero geometric albedos retrieved with this assumption) leads to a similar but slightly lower-skewed distribution of $A_{g}=0.40^{+0.29}_{-0.27}$ (Appendix Figure \ref{fig:populationR1}). We now discuss potential interpretations of this dichotomy.

\subsubsection{Selection/Detection Bias}
\begin{figure*}
    \centering
    \includegraphics[width=1.0\linewidth]{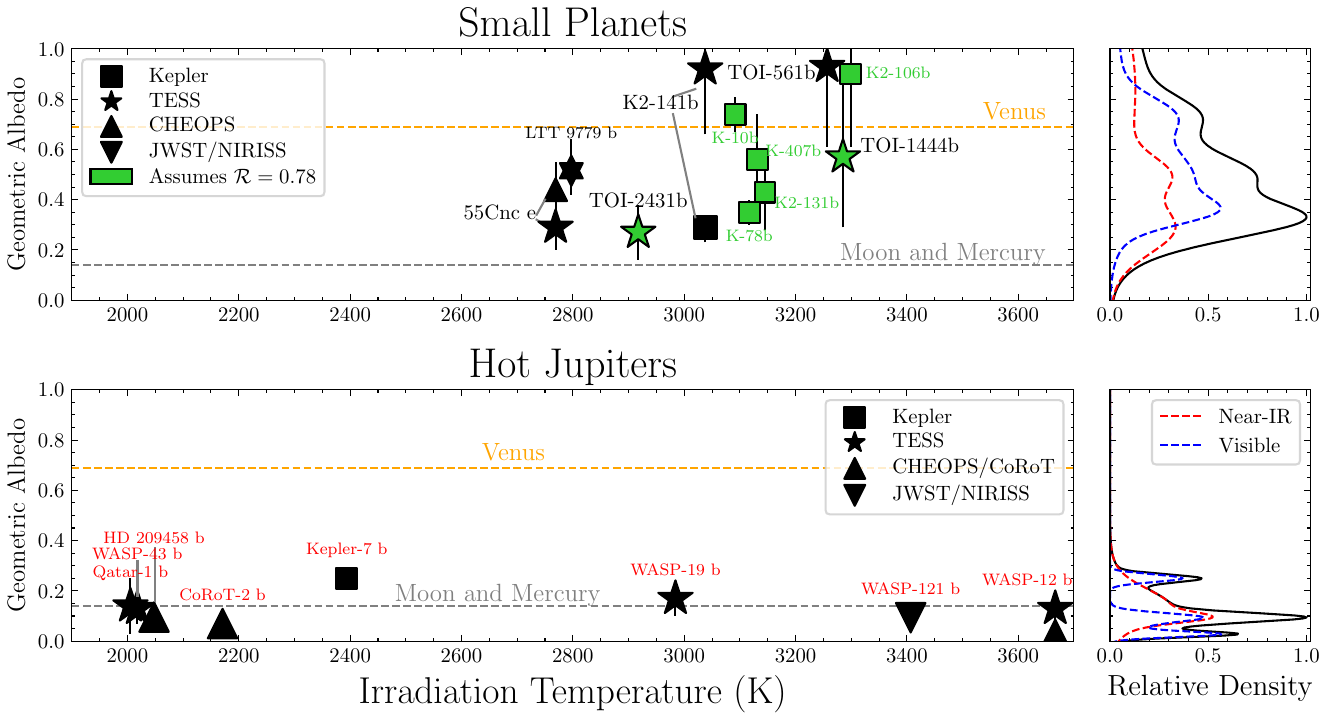}
    \caption{(top) Our homogeneously-derived geometric albedos for small ($<1.9\,R_{\oplus}$, plus LTT~9779~b) planets, compared to (bottom) select values derived for hot Jupiters compiled in \citet{heng2026albedo}.  Note that the NIRISS-($0.50\pm0.07$) and CHEOPS-derived ($0.53\pm0.11$) geometric albedos for LTT~9779~b overlap. Green points assume an dayside temperature ratio of $\mathcal{R}=0.78$, whereas black points are based on temperature constraints from mid-infrared observations (see Section \ref{sec:r78}). The values for small planets (barring LTT~9779~b NIRISS) do not take into account contribution from atmospheric emission, and thus may overestimate the true geometric albedo (Section \ref{sec:planck}). (right) shows the combined distribution of geometric albedo values, assuming split Gaussian distributions for each measurement, showing that the two populations show clearly different behaviors. The distributions are further split to visible (Kepler/CHEOPS/CoRoT, blue) and near-infrared (TESS/NIRISS, red) observations, which generally show good agreement.}
    \label{fig:population}
\end{figure*}

An issue with the above comparison is that small planets carry an inherent observational bias; it may be the case that \textit{only} the planets with high albedos have detectable dayside fluxes.  This could imply that low albedo lava worlds, however numerous, are simply not detectable, and only focusing on detections biases the observed geometric albedo distribution.  The reflected light signal is strongly dependent on the planet's semi-major axis, and thus temperature, meaning that very few planets with $T_{irr}\lesssim2500$\,K have signals large enough to be detectable with TESS or Kepler.  This is not as much of a concern for hot Jupiters, which have significant thermal emission signals in the TESS/Kepler bandpasses.  However, \textit{no} hot Jupiter shown in Figure \ref{fig:population} shows a high geometric albedo $>0.30$ whereas several small planets confidently cross this threshold, indicating that the dichotomy between geometric albedos is likely physical and not a detection bias.

\subsection{Silicate Clouds}
Observations of hot Jupiters with JWST MIRI LRS show evidence for silicate clouds via the Si-O stretch feature (e.g., \citealt{grant2023jwst,inglis2024quartz}).  As the atmospheres of lava worlds are expected to be silicate-rich (e.g., \citealt{schaefer2009chemistry,kite2016atmosphere}), reflective silicate clouds may explain their high geometric albedos.  \citet{coulombe2025highly} showed that the reflected light spectrum of the ultrahot-Neptune LTT~9779~b is most consistent with silicate clouds made of either Mg$_2$SiO$_4$ or MgSiO$_3$.  However, the spectrum did not show spectral features, so the exact cloud composition is uncertain.

Several 1D radiative-convective models of silicate vapor atmospheres show strong thermal inversions (e.g., \citealt{zilinskas2022observability,piette2023rocky}).  In 1D, inversions stop clouds from forming, as there is no region of the atmosphere that allows the vapor to condense (i.e., the temperature-pressure profile never falls below the condensation curve). However, clouds could still form as vapor moves away from the hot substellar point to colder regions \citep{nguyen2024clouds}.  The lower surface gravity of lava worlds compared to hot Jupiters may make it easier to loft small cloud particles high into the atmosphere where they can reflect incoming starlight more efficiently.

Future work might model how silicate clouds form in lava world atmospheres and their impact on observations, including radiative feedback and the impact of 3D atmospheric dynamics on clouds.  Predictions for the specific composition and particle size distribution of these clouds may be testable with JWST/NIRISS reflected light spectral observations (e.g., \citealt{coulombe2025highly}). More precise reflected light phase curves of lava worlds from JWST/NIRISS or future visible-band missions like PLATO may give more constraints on their longitudinal distribution.

\subsection{High Albedo Surfaces?}

Some laboratory work suggests that the emissivity of silicate material decreases at the high temperatures relevant to the lava worlds analyzed in this work (e.g., \citealt{thompson2021quantitative,fortin2024lava}).  Assuming these materials are opaque, this would imply that the reflectivity and thus albedo of these materials increase as they are melted.  If this were true, it could explain the dichotomy in the geometric albedos of lava worlds and hot Jupiters shown in Figure \ref{fig:population}.

These laboratory measurements, however, lack crucial wavelength coverage in the visible-near IR, which contains the majority of starlight relevant to energy balance and reflected light. The conversion from emissivity to reflectance also requires assuming that these samples are opaque, which has been questioned (e.g., \citealt{ferkl2026heat}). Future laboratory work measuring the \textit{direct reflectance} of silicate material at ultra-hot temperatures relevant to lava worlds will be required to definitively rule out surfaces as a cause for these planets' high albedos.  However, current data (e.g., \citealt{essack2020low}) points towards low surface albedos for these planets.

\section{Limitations and Future Work} \label{sec:futurework}

\subsection{Planck Assumption and Thermal Inversions}\label{sec:planck}

Work with TESS phase curves of hot Jupiters has shown that self-consistent atmospheric modeling is needed when deriving geometric albedos \citep{arora2024constraining}: assuming Planck function-like blackbody emission using temperatures derived from mid-infrared secondary eclipse observations causes over-estimates of geometric albedo. This is mostly due to atomic and molecular emission features from prominent thermal inversions increasing visible-band emission, but can also be caused by emission from deeper, hotter regions of the atmosphere.  One example is WASP-121~b, whose TESS-derived geometric albedo changes from $A_{g}\sim0.27$ to $0.00^{+0.06}_{-0.10}$ after modeling the wavelength dependence of the atmosphere's thermal emission contribution \citep{arora2024constraining}. Recent JWST/NIRISS observations, however, have since revised this value to $0.09\pm0.03$ \citep{splinter2025precise}.

A key issue in applying this methodology to lava worlds is that their atmospheric compositions are unknown, whereas hot Jupiter spectra are well-modeled assuming a near-solar composition gas mixture. Some models of lava world atmospheres assume that any initial volatiles, including H$_2$O, CO$_2$, or CO, are lost to space, resulting in atmospheric compositions set by silicate vapor pressure equilibrium (e.g., \citealt{kite2016atmosphere,zilinskas2022observability}).  However, other models are more agnostic, and allow for volatiles like H$_2$O, CO$_2$, or CO (e.g, \citealt{piette2023rocky,van2025lavatmos,zilinskas2025characterising}).  In addition, JWST NIRCam observations seem to support CO (or CO$_2$?) in the atmosphere of 55~Cancri~e \citep{hu2024secondary,patel2024jwst,snellen2026strong}.

\begin{figure*}
    \centering
    \includegraphics[width=0.9\linewidth]{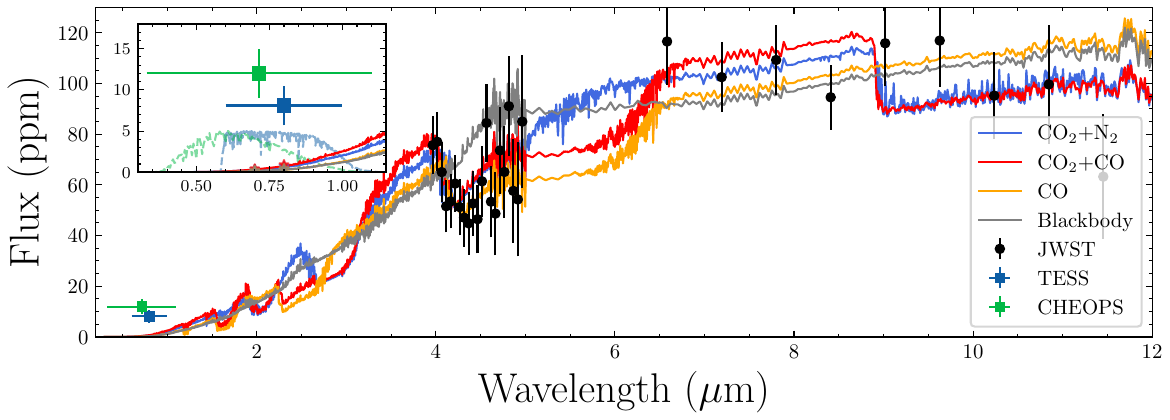}
    \caption{\textbf{Atmospheric models of 55~Cancri~e suggest that the optical TESS and CHEOPS dayside fluxes imply a reflective component unexplainable by thermal emission alone.}  Here, we show the TESS and CHEOPS-derived dayside fluxes from our analysis and \citet{demory202355} compared to the combined JWST NIRCam + MIRI LRS data presented in \citet{hu2024secondary}. The TESS and CHEOPS throughput functions, convolved with a PHOENIX stellar model, are shown as dashed lines. We include best-fit atmospheric models from \citet{hu2024secondary} for CO$_2$+N$_2$ and CO$_2$+CO mixtures, pure CO, as well as their best-fit blackbody model. Each model is expected to have $<1\,$ppm contribution in the TESS and CHEOPS bandpasses.  While the atmospheric models indeed show greater emission in the visible than the simple blackbody assumption, no model can adequately explain the dayside fluxes from CHEOPS and TESS, implying a reflective component.}
    \label{fig:55cncspec}
\end{figure*}

For 55~Cancri~e, however, the fact that reported carbon (CO or CO$_2$) features seem to be in absorption rather than emission, suggesting a lack of prominent thermal inversions, can help constrain this degeneracy. We show 55~Cancri~e's visible dayside fluxes with CHEOPS and TESS compared to JWST MIRI LRS and NIRCam observations and best-fit models presented in \citet{hu2024secondary} in Figure \ref{fig:55cncspec}. Each atmospheric model is expected to have $<0.9\,$ ppm contribution in the TESS bandpass and $<0.4\,$ppm contribution in the CHEOPS bandpass.  The maximum difference between the bandpass-integrated atmospheric and blackbody models is 0.5\,ppm and 0.2\,ppm in TESS and CHEOPS, respectively, biasing the geometric albedo measurement by--at most--0.02 or 0.01, respectively. While the atmospheric models do indeed show greater thermal emission in the visible bandpass than the Planck blackbody assumption, the dayside fluxes are unexplainable with atmospheric emission alone, implying an additional reflective component.

While we focus on this small subset of models, the effective surface pressure of lava world atmospheres, which sets how hot the deep atmosphere can be, is unknown. Moreover, it is unclear whether/how silicate clouds would mute potential absorption and emission features. Future mid-IR observations of lava worlds to better characterize their atmospheric composition and thermal structure will be needed to break these degeneracies.

\subsection{Assumption of Negligible Eccentricity}
So far we have assumed perfectly circular orbits, so that the mid-eclipse occurs at 0.5 orbital phase.  Recent JWST secondary eclipse observations of USP lava worlds indicate very low eccentricities $\lesssim0.01$ (JWST GO 4818, priv. comm.), however this may not be the case for all planets in our sample.  Eccentricity affects the phase curve both through the mid-eclipse timing ($t_{secondary}$) and the eclipse duration ($T_{14,ecl}$) compared to the transit duration ($T_{14,trans}$) \citep{winn2010exoplanet}, where  

\begin{equation}
    \Delta t_{secondary}\approx \frac{2P}{\pi}\times e\cos\omega,
\end{equation}

and,
\begin{equation}
    \frac{T_{14,ecl}}{T_{14,trans}}\approx \frac{1+e\sin\omega}{1-e\sin\omega}.
\end{equation}
Note that $e\sin\omega$ is generally much more difficult to constrain than $e\cos\omega$ for low-SNR eclipse observations.

Allowing for free eccentricity can lead to the spurious detection of false positive eclipses if random systematics line up to create an eclipse-like event, especially if the eclipse duration is short compared to the orbital period. For example, consider TOI-6255~b, an Earth-sized USP planet ($R=1.08\,R_{\oplus}$, $P=0.238$\,d) orbiting a nearby M dwarf.  \citet{dai2024earth} analyzed the TESS sector 16 and 56 phase curve of the planet, retrieving a dayside flux of $43\pm25$\,ppm after enforcing zero nightside flux, compared to the maximum reflected signal of 74\,ppm and maximum thermal signal of 4\,ppm. Adding new data from sector 83 and masking out the eclipsing binary TOI-6255.02, we find a similarly high dayside flux of $42^{+27}_{-24}$~ppm.   However, the residuals show a clear eclipse-like shape slightly offset from 0.5 orbital phase.

We find that allowing for a free eccentricity fits an eclipse at $0.520^{+0.006}_{-0.004}$ phase, corresponding to $e\cos\omega=0.031^{+0.009}_{-0.006}$, with a deep eclipse depth of $101\pm32$\,ppm (3.1\,$\sigma$ from zero).  The nightside flux is also high at $76^{+31}_{-32}\,$ppm (2.4$\,\sigma$ from zero).  The high dayside flux, if real, corresponds to either a extremely high geometric albedo of $\sim1.4\pm0.5$ or an emitting temperature $1.4\times$ the theoretically expected maximum.  This would require an additional energy source $>$2$\times$ that of the dayside-averaged instellation. \citet{dai2024earth} estimates the tidal circularization timescale of the planet at $\sim400$ yr, so the non-zero eccentricity is  unlikely unless the planet's orbit is constantly perturbed by an (yet undetected) outer companion.

Assuming $e\sin\omega\sim e\cos\omega\rightarrow e\sim\sqrt2 \times e\cos\omega$, we can estimate an implied tidal heating rate assuming an Earth-like interior following the `fixed $Q$' approach of \citet{driscoll2015tidal,coy2025population} of $\sim 130\,$kW/m$^{2}$ assuming an Earth-like interior ($Q=100,k_{2}=0.3$). This value is comparable, but not greater than, the averaged dayside flux received from the star of $\sim405$\,kW/m$^{2}$. However, such a high tidal heating rate is unlikely due to negative feedback between internal temperature and dissipation rate \citep{driscoll2015tidal}. This also would result in a change in the brightness temperature ratio of only $\sim7\%$, and thus is not enough to explain the observation alone. The planet has been observed with a MIRI LRS phase curve (JWST GO 8864: PI Dang), which will be able to confirm or reject the potential non-zero eccentricity and extreme tidal heating.

\begin{figure*}
    \centering
    \includegraphics[width=0.95\linewidth]{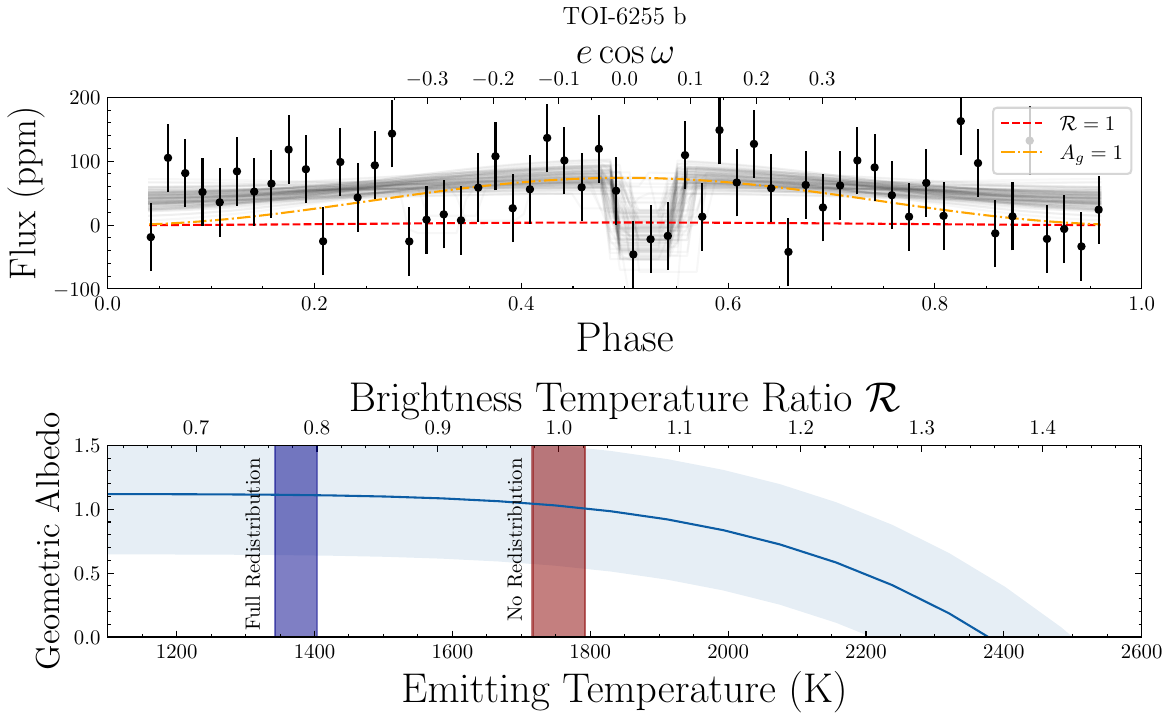}
    \caption{The sectors 16, 56, and 83 TESS phase curve of TOI-6255~b. The slightly-offset secondary eclipse phase of  $0.520^{+0.006}_{-0.004}$ suggests a potential non-zero eccentricity ($e\cos\omega=0.031^{+0.009}_{-0.006}$). The high dayside flux of  $101\pm32$\,ppm and nightside flux of $76^{+31}_{-32}\,$ppm, if real, suggest high amounts of internal heating. JWST MIRI LRS phase curve observations (JWST GO 8864: PI Dang) will be able to more strongly constrain a potential non-zero eccentricity.}
    \label{fig:ecosw}
\end{figure*}
\subsection{Planet Variability is Likely Undetectable in TESS}\label{sec:variability}
\begin{figure}
    \centering
    \includegraphics[width=0.9\linewidth]{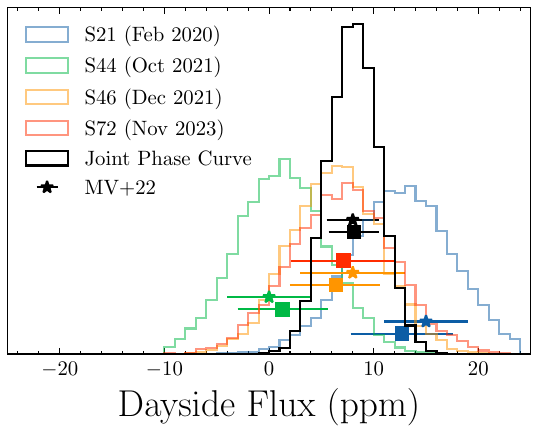}
    \caption{Retrieved dayside flux of 55~Cancri~e for each TESS sector compared to our full phase curve fit.  Stars represent values reported in \citet{valdes2022weak}, whereas squares and histograms represent our new analyses.  For the individual sectors, we use the same baseline prescription as \citet{valdes2022weak} (0.25--0.75 phase).  Overall, we find slightly less evidence for significant variability between sectors than \citet{valdes2022weak}.}
    \label{fig:variability}
\end{figure}

Planet variability is possible, though not emphasized in this study. 55~Cancri~e has been reported to have sub-week timescale variability in its near-infrared and optical eclipse depths \citep{valdes2023investigating,patel2024jwst,snellen2026strong}.  Variability might be due to cloud limit cycle feedbacks \citep{loftus2025extreme}, chaotic outgassing events \citep{heng2023transient}, or a circumstellar torus of evaporating material \citep{valdes2023investigating}.  Following \citet{valdes2022weak},  we analyzed TESS eclipses of 55~Cancri~e for variability between sectors using \texttt{SPARTESS}. We used the same baseline and phase curve shape (i.e., a flat phase curve) prescription as \citet{valdes2022weak}.  We used the sliding linear fit methodology for detrending the light curves.  We retrieve eclipse depths of $13\pm5$\,ppm, $1\pm4$\,ppm, $6\pm4$\,ppm, and $7\pm5$\,ppm for sectors 21, 44, 46, and 72, respectively, compared to $15\pm4$\,ppm, $0\pm4$\,ppm, and $8\pm5$ reported for sectors 21, 44, and 46 in \citet{valdes2022weak}.  We compare posterior distributions for each sector in Figure \ref{fig:variability}.

Overall, we also find weak evidence for variability between sectors, but less so than found in \citet{valdes2022weak}, and all individual sectors are 1.6$\,\sigma$ consistent with our retrieved value using the full phase curve. Thus, TESS likely lacks the precision to analyze visible/NIR variability of non-disintegrating lava worlds on subannual timescales.

\section{New Observatories: PLATO and Roman} \label{sec:futureobs}

\subsection{PLATO}\label{sec:plato}

The PLATO mission \citep{rauer2025plato}, set to launch in early 2027, should provide stringent constraints on the phase curves of hot Jupiters (e.g., \citealt{singh2021plato}).  The telescope will consist of 24 `Normal Cameras' (or `N-Cams') and two `Fast Cameras' (or `F-Cams') meant for brighter targets and pointing guidance. The N-Cams will observe in white light in a bandpass (500-1000\,nm) slightly bluer than TESS.  The cameras each have an aperture size of 12 cm (compared to TESS' 10 cm), resulting in an increase in effective area compared to TESS of $1.44\times N_{obs}$, where $N_{obs}$ is the number of cameras observing the target.

The mission should achieve photometric precision $\sim50-100\times$ better than TESS for bright stars covered by all 24 N-Cams and $\gtrsim3\times$ better than TESS for stars covered by at least 6 N-Cams \citep{rauer2025plato}.  PLATO plans to target two separate fields for 2 years each, each covering $\sim5\%$ of the sky.  This long stare observing strategy favors phase curve observations, allowing a potential $\sim160\times$ increase in photometric precision compared to 4 sectors (120 days) of TESS data. A list of known stars, including exoplanet hosts, within the first PLATO field has been published in \citet{montalto2026plato}.  These data include estimates for the host star's expected signal-to-noise and PLATO magnitude.   This list includes TOI-431~b and WASP-121~b, two planets (a lava world and an ultra-hot Jupiter analyzed above) with favorable observability metrics in the visible.  TOI-431~b will be observed by 6 N-Cams while WASP-121~b will be observed by 12.  Neither host star is expected to saturate the N-Cameras.  The expected noise per 25-second integration is only 350 ppm for TOI-431 and 530 ppm for WASP-121 based on the scaling reported in \citet{rauer2025plato}.

We simulated the expected precision of the phase curves of TOI-431~b and WASP-121~b using the end-to-end camera simulator \texttt{PLATOSim}. \texttt{PLATOSim} propagates a specified sky model through the PLATO optics and detector chain at the pixel level and extracts photometry using the on-board optimal aperture algorithm introduced in \citet{marchiori2019flight}.

The injected phase curves' transits and eclipses were modeled with \texttt{batman} \citep{kreidberg2015batman}; for TOI-431, we inject the transits of planet b and d alongside a visible phase curve signal of 10 ppm, consistent with our upper limit of 10.9\,ppm in the TESS bandpass and corresponding to a geometric albedo of 0.4.  The injected phase curve also has an eastward phase offset of 20$\degree$.  Orbital parameters were based on the median values reported in \citet{Osborn2021}, although Gaussian priors were used in lightcurve fitting. For WASP-121, we injected the transits alongside a visible phase curve with a 400 ppm dayside flux, 2 ppm nightside flux, 18 ppm ellipsoidal variation signal, and a 1.8 ppm Doppler boosting signal based on the theoretical values reported in \citet{wong2020systematic}.  The planet parameters were defined based on median values presented in \citet{sing2024absolute}.

We included stellar variability from rotational modulation by evolving, rigidly rotating starspots computed with the analytic model of \citet{kipping2012analytic}. The rotational modulation was rescaled to an assumed amplitude: $\sim8000\,$ppm peak-to-peak at $P_{\rm rot}=30.1\,$d for TOI-431, and $\sim500\,$ppm at $P_{\rm rot}=1.07\,$d for WASP-121.  We also injected granulation and solar-like oscillations using \texttt{PLATOSim}'s asteroseismic scaling relations: granulation as the two-component super-Lorentzian of \citet{kallinger2014connection}, and p-modes as solar frequencies rescaled to each target's $\nu_{max}$ (the frequency of maximum oscillation power) and large separation, with \citet{corsaro2013bayesian} mode amplitudes, realized as stochastically excited damped oscillators \citep{de2006discovery}.  TOI-431 receives 26\,ppm RMS of granulation and 4.5\,ppm RMS of oscillations. The oscillations are centered on $\nu_{max}=5002\,\mu$Hz. WASP-121, a larger and hotter star, receives stronger signals of 51 and 12 ppm RMS, respectively, centered on a lower $\nu_{max}$ of 1897 $\mu$Hz (~9 min).

\begin{figure*}
    \centering
    \includegraphics[width=0.95\linewidth]{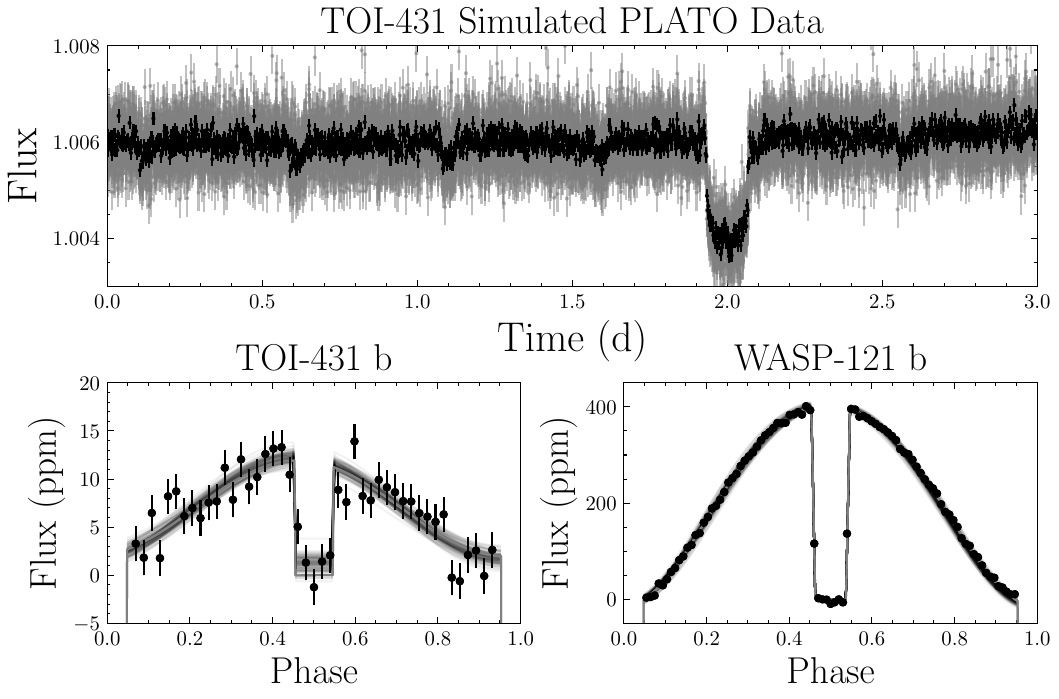}
    \caption{Simulations of PLATO photometric data for TOI-431 and WASP-121 using \texttt{PLATOSim} \citep{jannsen2024platosim}.  Both targets will be observed by the first PLATO field with 6 and 12 NCams, respectively.  The top panel shows simulated data for TOI-431, including the effects of spot-induced rotational modulation, granulation, and oscillation noise (see Section \ref{sec:plato}).  Grey data represents the raw 25 s cadence whereas black data is binned to the TESS cadence of 2 min. Using the full 2-year baseline, we recover a 10\,ppm signal for TOI-431~b at $\sim11\,\sigma$ ($10.9\pm1.0$\,ppm) and a small phase offset of 20$\degree$ ($15\pm4$$\degree$).  We also recover the WASP-121~b dayside flux at $105\,\sigma$ confidence ($420\pm4$\,ppm). These results show that PLATO should have the sensitivity to characterize small planet phase curves.}
    \label{fig:simplato}
\end{figure*}

For TOI-431, we used the sliding linear fit method to detrend the lightcurve, which resulted in no evidence for additional correlated noise, and removed transits of planet d. We retrieve a dayside flux of $10.9\pm1.0$\,ppm, nightside flux of $0.7\pm1.0$\,ppm, and phase offset of $15\pm4$ degrees east. Simulated data and the full phase curve are shown in Figure \ref{fig:simplato}.  The resulting RMS of the residuals is 406\,ppm, 16\% higher than the 350\,ppm predicted by the scaling of \citet{rauer2025plato}.  All retrieved phase curve values are close to the injected values, highlighting that PLATO should have the photometric precision to characterize very small signals similar to that achieved for Kepler.  

For WASP-121~b, we used a GP with the M\'atern-3/2 kernel to detrend the lightcurve following our TESS analysis.  We retrieve a dayside flux of $420\pm4$\,ppm (a 105$\,\sigma$ detection), phase offset of $3\pm2$$\degree$, and a nightside flux of $15\pm16$\,ppm.  The dayside flux is slightly higher (20\,ppm, $5\,\sigma$) than the injected value, likely due to using a single GP model despite the spot properties evolving over time. Real PLATO data will likely require splitting the systematics models into smaller sectors instead of using one global model, as done here. The increased precision of PLATO may allow for the direct granulation and oscillation noise of the star to be empirically constrained and modeled out of the lightcurve. These results highlight that PLATO is slated to place tight constraints on the phase curves of ultra-hot Jupiters.

\subsection{Roman}\label{sec:roman}

The recently-launched Nancy Grace Roman Space Telescope will conduct a Galactic Bulge Time-Domain Survey (GBTDS) with six `high cadence seasons' relevant to exoplanet detection and characterization.  These seasons will involve continuous stares into the galactic bulge each of 72 days: three at the beginning of the mission, and three near its end. Each season will be split into six-tile `mosaics', with each mosaic region having one 66 second integration per 12.1 min cadence cycle. The data will be collected using the F146 broadband filter (0.93--2.00$\,\mu$m).  This wavelength range is redder than both TESS and Kepler, and thermal emission and reflection may both significantly contribute to the Roman phase curves.  The GBTDS is expected to detect several thousand planet candidates, including $\sim1000$ planets smaller than 2$\,R_{\oplus}$ \citep{wilson2023transiting}.  However, given the expected dimness ($\mathbf{J}\gtrsim15.5$) of the host stars, follow-up observations will be difficult. Phase curve information could be a useful additional constraint from the GBTDS.

We use the Roman ETC\footnote{\url{roman.etc.stsci.edu}} to simulate the expected signal-to-noise of transiting exoplanet phase curve observations with the GBTDS.  The maximum SNR per integration achievable for partial saturation is roughly equal to the maximum SNR of no saturation, meaning that brighter stars will not result in significantly more precise phase curves. In addition, bright sources observed in the GBTDS are expected to produce complex persistence effects and bleed significantly into nearby pixels \citep{2026arXiv260718419L}. 

We focus on two end-member cases: a K2-141~b-like lava world (bandpass-integrated signal size 100 ppm), and a WASP-121~b-like ultra hot Jupiter (signal size 800 ppm). For our simulations, we assume that the planets orbit a star right below the saturation limit ($\mathbf{J}\sim17.3$), where the SNR is 470 per integration.  We inflate the expected error by 16\% based on the \texttt{PLATOSim} results above, resulting in 2468\,ppm precision per 12.1 min duty cycle.  This is $\sim25\times$ worse than the 98\,ppm precision expected for PLATO observations of WASP-121 b and $5\times$ worse than the $\sim473$\,ppm precision achieved by TESS for the same cadence, primarily due to the dimness of the injected host star.  We inject rotation-induced variability via the same methodology used in our PLATO simulations, where K2-141 has a peak-to-trough variability signal of 1\% and a rotation period of 14.03 days.

\begin{figure}
    \centering
    \includegraphics[width=0.95\linewidth]{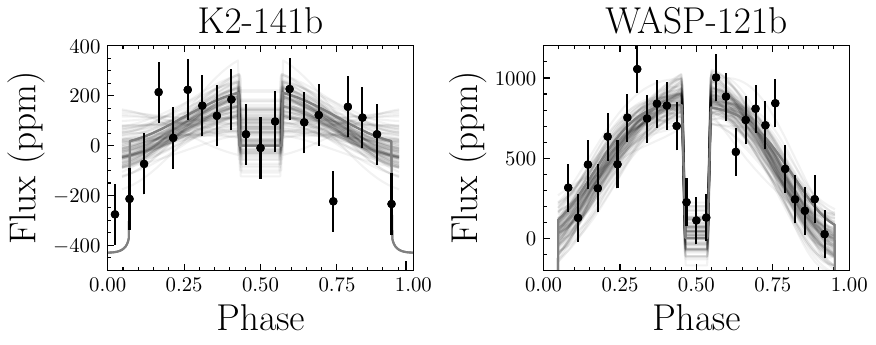}
    \caption{Simulated 72-day (one season of the Galactic Bulge Time Domain Survey/GBTDS) phase-folded phase curves of a K2-141~b-like lava world and WASP-121~b-like hot Jupiter using SNR estimates from the Roman ETC (Section \ref{sec:roman}).  While the transit signal for K2-141~b is detected at $6\,\sigma$ above zero, the dayside flux is not detected. The WASP-121~b analogue's dayside flux is detected at $\sim10\,\sigma$.  These results highlight that the Roman GBTDS will likely not provide useful constraints for small planet phase curves, but may provide useful constraints for planets with much larger signal sizes (i.e., ultra-hot Jupiters).}
    \label{fig:simroman}
\end{figure}

We do not consider the potential effects of dilution and charge migration from nearby saturated pixels \citep{2026arXiv260718419L} that will likely affect the real data and worsen SNR. \textit{The phase curves shown in Figure \ref{fig:simroman} thus represent the maximum one-season precision achievable under absolutely ideal conditions.}  

We use the sliding linear fit to detrend the K2-141 data and use the M\'atern-3/2 kernel to detrend the WASP-121 data due to the close stellar rotational period.
Retrievals of the simulated data using loose Gaussian priors for orbital parameters result in a $\sim10\,\sigma$ detection of the dayside flux for the WASP-121~b analogue ($894^{+71}_{-91}$)\,ppm and only a $\sim1\,\sigma$ detection for the K2-141~b analogue ($119^{+102}_{-104}$\,ppm). Notably, the transit signal for K2-141~b is only detected at $\sim6\,\sigma$ above zero.  The dayside flux precision is low, so even adding 5 new seasons (resulting in a $\sim2.4\times$ increase in precision) observing the same star would not result in a $>3\,\sigma$ detection.

As these are optimistic simulations, we conclude that Roman will likely not give high-quality phase curve constraints for small planets.  However, Roman may help constrain the phase curves of ultra-hot Jupiters, which have larger signal sizes.

\section{Conclusions} \label{sec:conclusions}
We present \texttt{SPARTESS}, a new pipeline for analyzing exoplanet phase curves with TESS data.  Our pipeline shows good agreement in the retrieved dayside fluxes for the gas giants LTT~9779~b and WASP-121~b when compared to JWST NIRISS phase curve observations.  We use \texttt{SPARTESS} to uncover new detections of the TESS (0.6-1.0$\,\mu$m) dayside fluxes of the lava worlds TOI-2431~b, K2-141~b, 55 Cancri e, TOI-561 b, and TOI-1444~b, alongside the first $>3\,\sigma$ dayside flux detection for LTT~9779~b. The four planets that have mid-IR constraints on their dayside temperatures (LTT~9779~b, K2-141~b, 55~Cancri~e, and TOI-561~b) have dayside fluxes much higher than expected for thermal emission alone, implying we are directly detecting the planet's reflection of the host star's light. We introduce an observability metric for investigating the reflected light signals of small planets, the Reflection Spectroscopy Metric (RSM), and show that our analyses of TESS data are near photon-limited. Our results show that the moderate-to-high geometric albedos of lava worlds in the visible suggested by Kepler extends to the redder near-infrared wavelength range probed by TESS.  These results support an emerging picture of highly reflective, likely silicate clouds being common on lava worlds, although laboratory measurements of the reflectivity of lavas at the extreme temperatures of these planets ($\sim2000-3500\,$K) will be required to rule out reflective lava surfaces.  If these planets do indeed host silicate clouds, they show stark contrast with hot Jupiters, which (at similar temperatures) are dark.  Future modeling will be required to investigate the origin of these high geometric albedos on lava worlds.   The upcoming PLATO mission is expected to enable stringent constraints on the phase curves of lava worlds.

\begin{acknowledgments}
This research has made use of the NASA Exoplanet Archive, which is operated by the California Institute of Technology, under contract with NASA under the Exoplanet Exploration Program. This paper includes data collected by the TESS mission. All the TESS data used in this paper can be found in MAST: \dataset[10.17909/t9-wpz1-8s54]{http://dx.doi.org/10.17909/t9-wpz1-8s54}. This work was completed in part with resources provided by the University of Chicago's Research Computing Center.

The authors thank Edwin Kite, Eliza M.-R. Kempton, and Michael Radica for helpful comments in improving the manuscript. We also thank Natalie Batalha, Jonathan Fortney, Avi Sphorer, and Lisa Dang for insightful discussions.  M.Z. acknowledges support from the Heising-Simons Foundation through the 51 Pegasi b Fellowship Program.

\end{acknowledgments}

\facilities{TESS, JWST, CHEOPS, Kepler, Spitzer, Roman, PLATO}



\appendix
\restartappendixnumbering

\section{Full Phase Curve Fit Results}

\begin{figure*}
    \centering
    \includegraphics[width=0.9\linewidth]{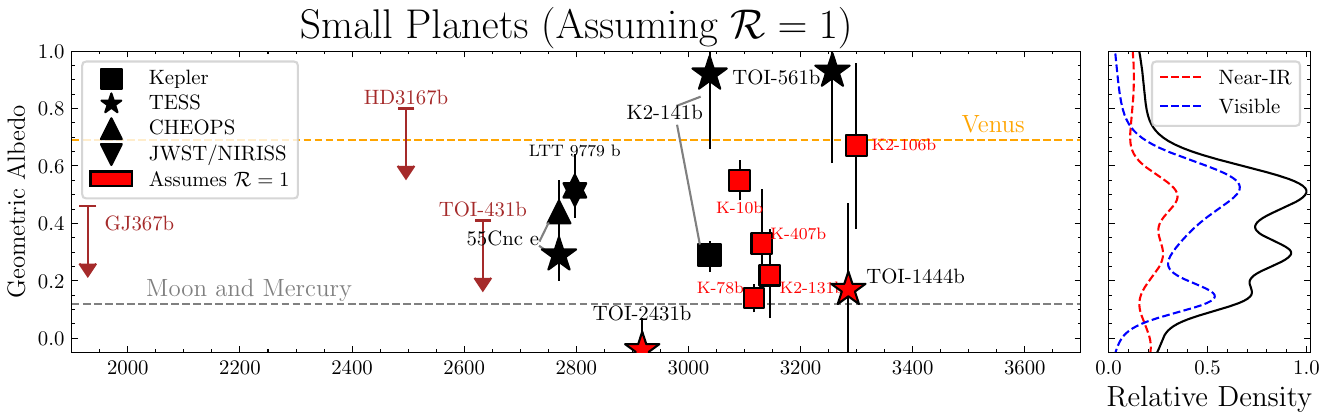}
    \caption{\textbf{Even energetically unrealistic lower bounds still recover the core result that small planets have higher albedos than hot Jupiters.} Here, when deriving geometric albedos for planets without mid-infrared dayside temperature constraints, we assume maximally hot daysides ($\mathcal{R}=1$).  This assumption is over-pessimistic in that it assumes that the planet reflects no light, yet the moderate-to-high derived geometric albedos suggest that a significant portion of light is being reflected.  We also include $2\,\sigma$ (97.7\%) upper limits on TESS geometric albedos derived from our analyses, and the $3\,\sigma$ upper limit for Kepler for HD~3167~b derived from \citet{vanderburg2016two,Coy2026}.}
    \label{fig:populationR1}
\end{figure*}

In Table \ref{tab:lttwasp} we show the full priors and posteriors used in our phase curve fits for LTT~9779~b, WASP-121~b, 55~Cancri~e, TOI-2431~b, K2-141~b, TOI-561~b, and TOI-1444~b.

\begin{deluxetable*}{lcccccccccc}
\tabletypesize{\scriptsize}
\tablenum{A1}
\tablecaption{TESS Phase Curve Parameters Derived for LTT~9779~b, WASP-121~b, 55~Cancri~e, TOI-2431~b, K2-141~b, TOI-561~b, and TOI-1444~b}\label{tab:lttwasp}
\tablewidth{0pt}
\tablehead{
\colhead{Planet} & \colhead{Dayside Flux} & \colhead{Offset$^{1}$} & \colhead{Nightside Flux} & \colhead{Radius} & \colhead{Semi-Major Axis} & \colhead{Inclination} & \colhead{Period} & \colhead{Time of Conjunction ($t_0$)} & \colhead{$q_1$} &\colhead{$q_2$}\\
\colhead{} & \colhead{(ppm)} & \colhead{degrees east} & \colhead{(ppm)} & \colhead{($R_{p}/R_{\star}$)} & \colhead{($a/R_{\star}$)} & (deg) & \colhead{(days)} & \colhead{(BJD)} & \colhead{} &\colhead{}
}
\startdata
LTT 9779 b\\
Priors & $\mathcal{U}(-1000,1000)$ & ... & $\mathcal{U}(-1000,1000)$ &$\mathcal{U}(0.04,0.05)$ & $\mathcal{N}(3.877,0.091)$ & $\mathcal{N}(76.4,0.4)$ & $\mathcal{N}(0.79206410,1.4e-7)$ & $\mathcal{N}(2459043.31060,0.00009)$ & $\mathcal{U}(0,1)$ & $\mathcal{U}(0,1)$  \\
 Posteriors & $80.2\pm15.9$ & $5.3^{+9.8}_{-9.6}$ & $35.3^{+16.0}_{-15.6}$ & $ 0.0462^{+0.0020}_{-0.0017}$ & $3.86\pm0.06$ & $76.3\pm0.3$ & $0.7920641\pm1e-7$ & $2459043.31062\pm0.00007$ & $0.39^{+0.23}_{-0.25}$ & $0.36^{+0.30}_{-0.25}$ \\ \hline
WASP-121~b\\
Priors & $\mathcal{U}(-1000,1000)$ & ... & $\mathcal{U}(-1000,1000)$ &$\mathcal{U}(0.10,0.15)$ & $\mathcal{N}(3.7844,0.0069)$ & $\mathcal{N}(87.6,1.6)$ & $\mathcal{N}(1.274924762,4.6e-8)$ & $\mathcal{N}(2458661.563783,3.0e-5)$ & $\mathcal{U}(0,1)$ & $\mathcal{U}(0,1)$ \\
Posteriors & $410.0^{+18.1}_{-18.7}$ & $2.1^{+5.1}_{-5.0}$ & $31.7^{+36.5}_{-37.2}$ & $0.12175^{+0.00018}_{-0.00017}$ & $3.7838\pm0.0061$ & $87.6\pm0.2$ & $1.27492478^{+2.6e-8}_{-2.5e-8}$ & $2458661.563779\pm2.3e-5$ & $0.190^{+0.019}_{-0.018}$ & $0.283^{+0.034}_{-0.032}$  \\ \hline
TOI-2431~b\\
Priors & $\mathcal{U}(-1000,1000)$ & ... & $\mathcal{U}(-1000,1000)$ &$\mathcal{U}(0.01,0.03)$ & $\mathcal{N}(2.083,0.030)$ & $\mathcal{N}(74.00,0.99)$ & $\mathcal{N}(0.22419577,5e-8)$ & $\mathcal{N}( 2460258.86855,1.5e-4)$ & $\mathcal{U}(0,1)$ & $\mathcal{U}(0,1)$ \\
Posteriors & $35.3^{+10.9}_{-11.2}$ & $-19.6_{-16.8}^{+14.4}$ & $2.5^{+10.8}_{-10.5}$ & $0.0209\pm0.0003$ & $2.075\pm0.022$ & $74.35\pm0.69$ & $0.22419577\pm4e-8$ & $2460258.8685\pm0.0001$ & $0.57^{+0.21}_{-0.18}$ & $0.20^{+0.26}_{-0.14}$  \\ \hline
K2-141~b\\ 
Priors & $\mathcal{U}(-1000,1000)$ & ... & $\mathcal{U}(-1000,1000)$ &$\mathcal{U}(0.01,0.03)$ & $\mathcal{N}(2.366,0.047)$ & $\mathcal{N}(86.2,2.7)$ & $\mathcal{N}(0.280324956,6.7e-8)$ & $\mathcal{N}(  2457744.0717,2e-4)$ & $\mathcal{U}(0,1)$ & $\mathcal{U}(0,1)$ \\
Posteriors & $79.7^{+19.9}_{-19.8}$ & $14.1^{+11.6}_{-11.0}$  & $-1.6^{+20.1}_{-20.3}$ & $0.0199^{+0.0005}_{-0.0006}$ & $2.381^{+0.027}_{-0.037}$ & $86.0\pm2.1$ &  & $2457744.07174\pm0.00017$ & $0.22^{+0.21}_{-0.12}$ & $0.51\pm0.32$  \\ \hline
55~Cancri~e\\
Priors & $\mathcal{U}(-1000,1000)$ & ... & $\mathcal{U}(-1000,1000)$ &$\mathcal{U}(0.01,0.03)$ & $\mathcal{N}(3.52,0.01)$ & $\mathcal{N}(83.6,0.5)$ & $\mathcal{N}(0.73654625,1.5e-7)$ & $\mathcal{N}(  2459370.807543,9.3e-5)$ & $\mathcal{U}(0,1)$ & $\mathcal{U}(0,1)$ \\
Posteriors & $8.1\pm2.4$ & $-24.8^{+12.9}_{-14.0}$  & $1.2^{+2.4}_{-2.3}$ & $0.01805\pm0.0001$ & $3.52\pm0.01$ & $84.0\pm0.2$ & $0.73654631\pm9e-8$  & $2459370.807525\pm6e-5$ & $0.56^{+0.12}_{-0.10}$ & $0.13^{+0.09}_{-0.07}$  \\ \hline
TOI-561~b\\
Priors & $\mathcal{U}(-1000,1000)$ & ... & 0 (fixed) &$\mathcal{U}(0.01,0.03)$ & $\mathcal{N}(2.683,0.029)$ & $\mathcal{N}(87.0,2.1)$ & $\mathcal{N}(0.44656985,7.8e-7)$ & $\mathcal{N}(  2458517.94528,7.2e-4)$ & $\mathcal{U}(0,1)$ & $\mathcal{U}(0,1)$ \\
Posteriors & $31.8^{+11.2}_{-10.9}$ & $13.9^{+16.2}_{-15.3}$  & $3.6\pm11.1$ & $0.0152\pm0.0004$ & $2.664\pm0.022$ & $87.8^{+1.3}_{-1.5}$   & $0.4465693\pm 2e-7$ & $2458517.94487^{+0.00047}_{-0.00045}$ & $0.20^{+0.23}_{-0.12}$ & $0.29^{+0.35}_{-0.21}$  \\ \hline
TOI-1444~b\\
Priors & $\mathcal{U}(-1000,1000)$ & ... &$\mathcal{U}(-1000,1000)$  &$\mathcal{U}(0.01,0.03)$ & $\mathcal{N}(2.730,0.078)$ & $\mathcal{N}(82,3)$ & $\mathcal{N}(0.4702743,1.1e-6)$ & $\mathcal{N}(  	2458711.3675,0.0011)$ & $\mathcal{U}(0,1)$ & $\mathcal{U}(0,1)$ \\
Posteriors & $17.7^{+7.3}_{-7.2}$ & $4.0^{+19.6}_{-18.7}$  & $-5.3^{+7.6}_{-7.3}$ & $0.01403\pm0.0003$ & $2.741\pm0.067$ & $82.1^{+1.7}_{-1.5}$   & $0.4702735\pm 2e-7$ & $2458711.3669^{+0.0003}_{-0.0004}$ & $0.21^{+0.21}_{-0.11}$ & $0.42^{+0.35}_{-0.29}$  \\
\enddata
\tablecomments{Priors for orbital and planet parameters are defined based on: \citet{edwards2023characterizing} for LTT~9779~b, \citet{sing2024absolute} for WASP-121~b, \citet{tacs2026earth} for TOI-2431~b, \citet{malavolta2018ultra,zieba2022k2} for K2-141~b, \citet{piotto2024architecture} for TOI-561~b, \citet{dai2021tks,polanski2024tess} for TOI-1444~b, and
\citet{bourrier201855,kokori2023exoclock} for 55~Cancri~e. $^{1}$This value is derived from fitting for $D_{1}$ (see Equation \ref{eq:phasecurve}) using a prior of $\mathcal{U}(-1000,1000)$. For WASP-121~b, we also fit for the second-order parameters $C_{2}$ and $D_{2}$ using wide uniform priors ($\mathcal{U}(-1000,1000)$), retrieving $C_{2}=19.7^{+10.0}_{-10.2}$\,ppm and $D_{2}=1.1^{+8.8}_{-9.0}$\,ppm.}
\end{deluxetable*}

\section{Injection Recovery Tests}\label{ap:injrec}

Our choice of detrending method may introduce biases in the retrieved phase curve parameters.  One possibility is a high nightside flux caused by an `overflattening' of the phase curve, as can clearly be seen for LTT~9779~b.  The severity of this bias may depend on the host star brightness and level of stellar activity.

Here, we include simple injection recovery tests on real TESS data of Sun-like stars without any known transiting exoplanets to assess these potential biases.  For these fits, we assume a planet radius of $1.7\,R_{\oplus}$, an equilibrium temperature of 2000\,K ($T_{irr}=2830\,$K), and an impact parameter of 0.3.  The orbital period, inclination, and semi-major axis are computed to be self-consistent with the host star properties.  For simplicity, the limb darkening coefficients are set to $q_{1}=0.4, q_{2}=0.3$ for each run.

\subsection{$\tau$ Ceti}
$\tau$ Ceti is a very bright ($m_{Gaia}=3.25$) nearby inactive G8.5V star with 3 sectors of TESS data.  We injected a phase curve with a dayside flux of 15\,ppm with a westward offset of 30$\degree$.  This corresponds to a geometric albedo of 0.54 (assuming negligible thermal emission). The injected planet has a period of 0.62\,d, inclination of 85.2$\degree$, and transit depth of 350\,ppm based on the host star properties.

While the star is quiet, the PDCSAP data show signs of correlated noise/discontinuities near momentum dump events in sectors 3 and 30. Our initial fit included this correlated noise, aiming to assess its effect on retrieved parameters. Using the sliding linear fit detrending, we retrieve a dayside flux of $14.6\pm1.3$, phase offset of $8.9\pm3.5\degree$, and nightside flux of $-0.5\pm1.3$\,ppm.  While the dayside and nightside fluxes are very near the injected values, the phase offset is $6\,\sigma$ lower than the injected value.

Our second fit excluded 5 regions of TESS data near these discontinuities.  We retrieve a dayside flux of $14.8\pm1.3$\,ppm, nightside flux of $0.8\pm1.3$\,ppm, and a phase offset of $24.6\pm3.6\degree$ (now 1.5$\,\sigma$ from the injected value).  These results highlight that while the dayside and nightside fluxes are seemingly not significantly impacted by correlated noise/large discontinuities, that the phase offset is extremely sensitive to it, and thus the phase offsets reported above should be interpreted with caution.

\subsection{GJ~3346}

GJ~3346 is a moderately active, bright ($m_{Gaia}=7.12$) K4V star with 3 sectors of TESS data. The TESS data are similar to K2-141---the star shows prominent rotational modulation on the order of 2.5\% for sector 6 and 1\% for sector 32 (stronger than the $\sim 0.6-1\%$ seen for K2-141).  Sector 5 also shows extreme levels of correlated noise in PDCSAP not present in SAP unlikely to be stellar in origin similar to K2-141's sector 42.

We injected a dayside flux of 30\,ppm with a westward offset of 30$\degree$, corresponding to a geometric albedo of 0.47. The injected planet has a period of 0.36\,d, inclination of 83.8$\degree$, and transit depth of 495\,ppm based on the host star properties.

We initially used a GP M\'atern-3/2 kernel for sector 5 while detrending sectors 6 and 32 using the sliding linear fit, all using PDCSAP fluxes.  This resulted in a dayside flux of $35\pm8$\,ppm, offset of $33^{+15}_{-13}$$\degree$ west, and nightside flux of $9^{+8}_{-9}$\,ppm, all consistent with the injected values. To assess the impact of sectors with high amounts of instrumental correlated noise, we ran an additional fit excluding sector 5. This fit retrieved a dayside flux of $33\pm9\,$ppm, nightside flux of $6\pm9\,$ppm, and an offset of $34^{+14}_{-13}$$\degree$ west. A fit of only sector 5 using a M\'atern-3/2 kernel results in uninformative constraints on the dayside and nightside ($F_{d}=51\pm15\,$ppm and $F_{n}=29\pm80\,$ppm). Taking the dayside flux at face value would result in a false positive detection of the dayside flux, although other diagnostics suggest the PDCSAP data for this sector should not be used.

We ran another fit following the methodology used for our K2-141~b, i.e. a SHO GP kernel and using SAP data for sector 5.  We retrieve a dayside flux of $25\pm7\,$ppm, nightside flux of $8\pm7\,$ppm, and a phase offset of $40^{+22}_{-17}$$\degree$ west, all consistent with the injected values.  This indicates that even in the face of strong rotation signals that small phase curves should be recoverable with our methodology.

\section{Comparison of RSM to Kepler Lava Worlds Observations}\label{ap:RSM_comparison}

\begin{figure}
    \centering
    \includegraphics[width=0.95\linewidth]{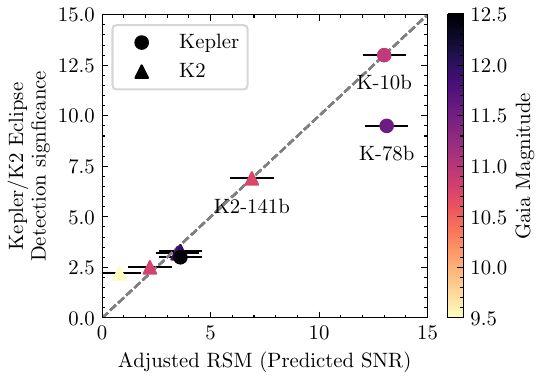}
    \caption{RSM versus the eclipse detection significance for the eight Kepler lava worlds reported in \citet{singh2022probing}.  The RSM has been adjusted to account for differences in observing time and the median retrieved signal size compared to the maximum reflected signal size (essentially $A_{g}$ is known a priori instead of assumed as in Section \ref{sec:RSM}).  Kepler targets (circles, Kepler-10~b, Kepler-78~b, and Kepler-407 b) are normalized to Kepler-10 b whereas K2 targets (triangles, K2-141~b, K2-131~b, K2-106~b, and K2-229~b, and K2-312~b) are normalized to K2-141~b. The dashed line indicates a one-to-one relation.}
    \label{fig:keplerrsm}
\end{figure}

To compared our SNR predictions (via RSM, see Section \ref{sec:RSM}) to observations, we compute an `adjusted RSM' which is meant to account for differences in the observing time (scaling RSM by a factor of $\sqrt{t/t_{ref}}$, where $t$ is the total observing time compared to a reference value $t_{ref}$ and assumes that the expected signal size is known a priori ($A_{g}$ used in the RSM calculation is unique per target based on their reported dayside fluxes).  The adjusted RSM is meant to estimate how well the actual noise follows the photon-noise scaling proposed by Equation \ref{eq:RSM}. We compare the adjusted RSM to the detection significance of 8 lava world dayside fluxes presented in \citet{singh2022probing}, finding a near one-to-one relation (Figure \ref{fig:keplerrsm}).  This shows that RSM is a good approximation for the signal-to-noise of reflected light observations if the expected $A_{g}$ is known a priori.  The large discrepancy between Kepler-78~b's adjusted RSM of 12.1 and real detection significance of 9.5 may be because the star is much more active than Kepler-10, and thus may have much larger uncertainties introduced during detrending. While we focus on photometry in this work, RSM also represents the relative signal-to-noise expected from spectroscopic observations. Very few known small planets have an RSM comparable to LTT~9779~b. Thus, LTT~9779~b's high geometric albedo could be a common (but infrequently detected) feature of small, hot planets.

\bibliography{main}{}
\bibliographystyle{aasjournalv7}



\end{document}